\documentclass[10pt,journal,compsoc]{IEEEtran}

\ifCLASSOPTIONcompsoc
  \usepackage[nocompress]{cite}
\else
  \usepackage{cite}
\fi

\ifCLASSINFOpdf
  \usepackage[pdftex]{graphicx}
  \graphicspath{{}}
 \DeclareGraphicsExtensions{.pdf,.jpeg,.png}
\else
\fi

\usepackage{amsmath}
\usepackage{amssymb}
\usepackage{array}

\usepackage{booktabs}
\usepackage{ragged2e} 
\usepackage{hyperref}

\begin{document}
%
\title{Estimating the Total Volume of Queries\\ to a Search Engine}
%
%
%
%

\author{Fabrizio~Lillo
        and~Salvatore~Ruggieri
\IEEEcompsocitemizethanks{
\IEEEcompsocthanksitem F. Lillo is with the Department of Mathematics, University of Bologna, Italy. 
E-mail: \href{mailto:fabrizio.lillo@unibo.it}{fabrizio.lillo@unibo.it}
\IEEEcompsocthanksitem S. Ruggieri is with the Department of Computer Science, University of Pisa, Italy. 
E-mail: \href{mailto:salvatore.ruggieri@unipi.it}{salvatore.ruggieri@unipi.it}
}}

%
%

\markboth{IEEE Transactions on Knowledge and Data Engineering}%
{}
%



\IEEEtitleabstractindextext{%
\justify
\begin{abstract}
We study the problem of estimating the total number of searches (volume) of queries in a specific domain, which were submitted to a search engine in a given time period. Our statistical model assumes that the distribution of searches follows a Zipf's law, and that the observed sample volumes are biased accordingly to three possible scenarios. 
These assumptions are consistent with empirical data, with keyword research practices, and with approximate algorithms used to take counts of query frequencies.
A few estimators of the parameters of the distribution are devised and experimented, based on the nature of the empirical/simulated data. For continuous data, we recommend using nonlinear least square regression (NLS) on the top-volume queries, where the bound on the volume is obtained from the well-known Clauset, Shalizi and Newman (CSN) estimation of power-law parameters. For binned data, we propose using a Chi-square minimization approach restricted to the top-volume queries, where the bound is obtained by the binned version of the CSN method. Estimations are then derived for the total number of queries and for the total volume of the population, including statistical error bounds. We apply the methods on the domain of \textit{recipes and cooking} queries searched in Italian in 2017.
The  observed volumes of sample queries are collected from Google Trends (continuous data) and SearchVolume (binned data). The estimated total number of queries and total volume are computed for the two cases, and the results are compared and discussed.
\end{abstract}

\begin{IEEEkeywords}
Search engine query, Volume estimation, Zipf's law, Power law, Google Trends.
\end{IEEEkeywords}}

\maketitle

\IEEEdisplaynontitleabstractindextext

%
\IEEEpeerreviewmaketitle

\IEEEraisesectionheading{\section{Introduction}\label{sec:introduction}}

%
%
%
%
\IEEEPARstart{T}{he} problem of computing the total number of searches (volume) of queries belonging to a specific domain is extremely relevant and, at the same time, challenging. For example, in web marketing the total volume ${\mathcal V}$ of queries quantifies the potential market of search engine advertising in the domain. When performing sociological or political research, instead, one might be interested in estimating the volume of queries belonging to a given macro-topic (e.g.,~immigration, ecology, health, etc.) searched in a given time period and from a given geographical area. In epidemiological investigations the total volume can be used to estimate the amount of information requested by the population for items related to some specific disease. An even more interesting quantity is the total volume ${\mathcal V}_v$ of queries searched at least $v$ times. In the web marketing example, ${\mathcal V}_v$  quantifies the potential market of queries with a minimum guaranteed volume. Related to the above, the total number of queries $N$ in the domain, or of queries $N_v$ searched at least $v$ times, are also extremely useful information for market analysis. 

However, the stream of queries submitted to a search engine is so massive that it is impractical to keep frequency counts of every possible query, particularly of those in the long tail of the distribution. 
In this paper, we study the problem of estimating the total volume of queries submitted to a search engine for a specific domain in a given time period. We design estimation methods from  sample data and experiment with them using both simulated and real data. As a specific example, we consider empirical data in the domain of  \textit{recipes and cooking}, which consists of queries with the name of the recipe of a dish, excluding drinks. The advantage over other domains is that it is relatively easy to collect sample recipes and to validate whether a given text is a recipe or not. In particular, we crawled popular websites of Italian recipes and cooking, collecting a sample of more than 120K distinct queries. We then resorted to Search Engine Optimization (SEO) tools for obtaining estimates of the number of searches to Google for each single query in the sample during the whole year 2017. We call such an estimate, the \textit{observed volume} of a query. We collected observed volumes from Google Trends\footnote{\href{https://trends.google.com}{\textit{https://trends.google.com}}} and SearchVolume\footnote{\href{https://searchvolume.io/}{\textit{https://searchvolume.io}}}. 

\begin{figure}[t]
	\centering
	\includegraphics[width=0.40\textwidth]{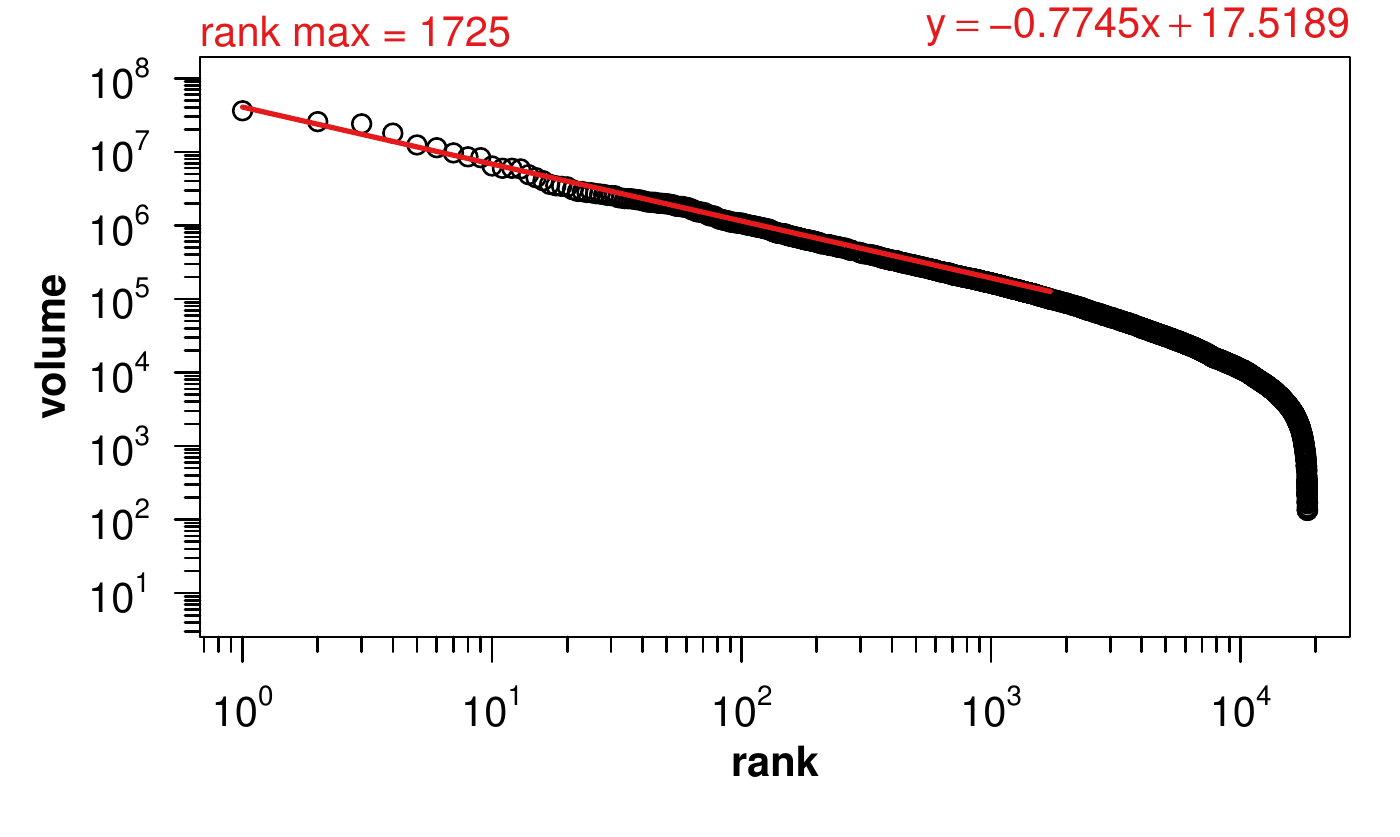}
	\caption{Empirical rank-volume distribution (scaled Google Trends estimates) for the investigated dataset (see Section \ref{sec:data} for details). Best view in color.
		\label{fig:empirical}}
\end{figure}
Our key problem is to estimate the total volume of the queries in the whole population, starting from  a possibly biased empirical sample of observed  volumes for a small set of queries. We start from the basic assumption that the rank-volume distribution of the whole population of queries (i.e. observed and unobserved) follows a Zipf's law. This assumption is supported by previous work \cite{DBLP:journals/tois/PetersenSL16} and by the empirical rank-volume distribution\footnote{The dataset is described in detail in Section \ref{sec:data}.} of Figure~\ref{fig:empirical}.

Observed volumes may be biased. Bias in observed volumes can be attributed to the adoption by SEO tools of sampling strategies and/or approximate counting techniques~\cite{DBLP:journals/ftdb/CormodeGHJ12}, e.g.,~\textit{count-min sketch} summaries \cite{DBLP:journals/jal/CormodeM05}. Such strategies favor volume estimation of popular queries against the ones in the long tail of the distribution. This yields the visible drop in volume in the tail of the empirical distribution of Figure~\ref{fig:empirical}. For instance, only 18.5K queries are assigned an observed volume estimate by Google Trends and only 12K queries by SearchVolume. Queries with low volume are monitored with lower probability by SEO tools, which then return no observed volume for them. We are able to model this behavior by assuming that the sample of observed volumes is not uniform, but it depends on the rank of queries over the population (\textit{non-uniform} sampling). Moreover, in order to account for approximations in the SEO tool data, we additionally assume that the observed volumes are noisy, and discuss two specific sampling schemes (\textit{noisy} and \textit{sketchy} sampling). 
The parameters of the Zipf distribution are estimated by a variant of Nonlinear Least Square (NLS) regression for continuous data, and of Chi-square optimization for binned data. 
Simulations show such estimators perform better than an alternative approach based on power law parameter estimation from continuous data \cite{DBLP:journals/siamrev/ClausetSN09} and binned data \cite{VirkarClauset2014} respectively. We derive then estimators of total volumes ${\mathcal V}$ and ${\mathcal V}_v$, and number of (distinct) queries $N$ and $N_v$, including  closed formula for statistical errors of such estimators.
In summary, this paper makes the following contributions:
\begin{itemize}
	\item we formalize the problem of estimating the total volume of queries submitted to a search engine, and propose a statistical model which is consistent with empirical data;
 	\item we design methods to infer parameters of the statistical model from both continuous and binned empirical data, and show that they perform well under simulated conditions;
	\item we apply the approach to the domain of recipes and cooking for queries in Italian, and produce estimations for the volume ${\mathcal V}_v$ of queries searched at least $v$ times in 2017 starting from empirical data collected from Google Trends (continuous data) and SearchVolume (binned data).
\end{itemize}

This paper is organized as follows. First, we report on related work in Section~\ref{sec:rw}. Next, we state the addressed problem in Section~\ref{sec:problem}. Section~\ref{sec:sampling} provides models of sampling bias. Section~\ref{sec:model} introduces and experiment with two estimators of the parameters of the Zipf's law, and builds on them for estimating the number and total volume of queries in the population.
Section~\ref{sec:binned} extends the approach to the case of binned empirical data. 
Section~\ref{sec:data} describes the empirical data obtained from Google Trends and SearchVolume tools, and it applies the estimators proposed in this paper to such data. Finally, conclusions summarize the contribution of the paper and open challenges for future work. 

\section{Related Work}\label{sec:rw}

Power laws distributions, Pareto distributions and Zipf's laws are ubiquitous in empirical data of many fields \cite{doi:10.1080/00107510500052444,DBLP:journals/siamrev/ClausetSN09}, and in information retrieval in particular  \cite{DBLP:journals/tois/PetersenSL16}.

Several works \cite{DBLP:journals/tois/PetersenSL16,DBLP:conf/wsdm/DingABS11,DBLP:conf/kdd/Baeza-YatesT07,DBLP:conf/sigir/Baeza-YatesGJMPS07} have observed that the probability that a query is searched $v$ times is approximately Power law distributed, namely $P(V=v) \propto 1/v^{\alpha}$. This information on  query frequencies has been used to optimize caching and distribution strategies in search engines and peer-to-peer systems. For our purposes, it implies  that the the probability that a query is ranked $i$-th follows a Zipf's law $P(R=i) \propto 1/i^{\beta}$ with $\beta = 1/(\alpha-1)$ (see e.g.,~\cite{DBLP:journals/jasis/Bookstein90,Adamic02}). Hence, we can resort to the  literature on the estimation of parameters of Power law distributions. Popular methods~\cite{DBLP:journals/tois/PetersenSL16} have relied on: graphical methods, straight-line approximation, maximum-likelihood estimation.  The estimated tail exponent, even in simulated data, significantly depends on the adopted method \cite{epjb}. A major breakthrough was the method proposed in~\cite{DBLP:journals/siamrev/ClausetSN09}, which consists in a maximum-likelihood estimation, with a cutoff for the fitting region determined with a Kolmogorov-Smirnov test. This method is implemented in the \textit{powerLaw}  package \cite{JSSv064i02} of  R \cite{Rlang}. 
Moreover, the method has been extended to the case of binned data in \cite{VirkarClauset2014}.

In the context of city size data, a standard approach for estimating the exponent of a Pareto distribution 
is to adopt ordinary (linear) least squares (OLS) regression of the log-ranks of cities based on their log-population \cite{NBERt0342}. Since Pareto fits well for the largest cities only,
\cite{RePEc:gla:glaewp:2012_10} proposes a recursive-truncation method (in the same line as \cite{DBLP:journals/siamrev/ClausetSN09}) for determining the best cut-off of minimum city size according to the Kolmogorov-Smirnov test.

The method developed in this paper makes the parametric assumption that the distribution of searches follows a Zipf's law. Empirical data investigated in this paper appear to agree with this assumption for more than three orders of magnitude (see Figure~\ref{fig:empirical}) and we attribute the deviations for very low rank queries to the sampling procedure rather than to a different functional form. An alternative explanation is that data in the population follows a different distribution (for example exponentially truncated Zipf's law). Indeed in other domains (e.g.,~city sizes) there is an ongoing debate about the accuracy of power law distribution in explaining empirical data (see \cite[Figure 1]{RePEc:gla:glaewp:2012_10} for cities and more generally \cite{Clauset2019,Holme2019,DBLP:journals/corr/abs-1811-02071}). We believe that our approach could be extended to functional forms different from exact Zipf's law, but this is clearly beyond the scope of the paper.

The \textit{unseen species} problem asks how many biological species are present in a region, given that in an observation campaign a certain number of species with their relative frequency have been observed. 
Despite there are several estimators for the unseen species problem (for example, see \cite{pnas2016}), the problem tackled here is different in an important aspect. In the unseen species problem, it is often assumed that in the sample used to build the estimator, the observed frequencies are proportional to the true frequencies in the population. In other words, there is no bias in the construction of the sample. In our approach, the elements of the sample are chosen {\it ex-ante} and the probability of being in the sample is not necessarily proportional to true frequency.

Regarding SEO tools, there is little documentation on how they collect, sample and filter query logs for providing observed volumes. The biases behind observed volumes reported by SEO tools remain unknown. Google Trends can rely on Google search engine logs. Independent SEO tools (Searchvolume, Ubersuggest, Semrush, Keywordkeg, etc.) rely on a more limited user base. \cite{DBLP:journals/jasis/VaughanC15} compares Google Trends and Baidu Index (restricted to searches from China only), and finds that their estimates are highly correlated. An advantage of Baidu Index over Google Trends is that it provides absolute estimates, not relative ones. 
Generally, SEO tools provide observed volumes of a given query, not aggregate volumes over a domain. One exception is Google Trends, which returns (relative) volumes for a topic, which is defined as ``a group of terms [a.k.a.,~queries] that share the same concept in any language". Such volumes are aggregated over the queries in the group monitored by Google Trends, which filters out low volume queries. Therefore, they are not estimates of the total volume of the population of queries, as considered in this paper.

Finally, this paper substantially extends preliminary results appeared in \cite{DBLP:conf/www/LilloR19}, which were restricted to the case of continuous data (Section~\ref{sec:model}). Moreover, \cite{DBLP:conf/www/LilloR19} largely focuses on the data collection issues, which are not part of this paper.

\section{Problem statement}\label{sec:problem}

Let us assume that the population of all (distinct) queries to a search engine in a reference domain and period of time is composed by $N$ queries. The volume of a query is the number of times the query is searched in the given period of time. We model the volumes as random variables and we label them in such a way that  $V_1 \ge V_2 \ge \ldots \ge V_N$.   In other words, when writing $V_i$, we intend the volume of $i$-th rank\footnote{These are also called order statistics.}.

We assume that the distribution of searches follows a Zipf's law, which predicts that the probability of observing a search of the query corresponding to $V_i$ is: 
\begin{equation}\label{zipf2}
f(i)=\frac{A}{i^\beta}
\end{equation}
where $A^{-1}=\sum_{i=1}^N i^{-\beta} = [\zeta(\beta)-\zeta(\beta, N+1)]$ is a normalizing constant and $\zeta(\beta) =  \sum_{i=1}^{\infty} i^{-\beta}$ and $\zeta(\beta, N+1) =  \sum_{i=N+1}^{\infty} i^{-\beta}$ are the Riemann zeta and Hurwitz functions, respectively.

If ${\cal V}$ is the total volume over the population, the expected volume of the query corresponding to $V_i$ is $\bar V_i={\mathcal V} f(i) =\frac{c}{i^\beta}$
where:
\begin{equation}\label{eq:area}
c=\frac{{\mathcal V}}{\zeta(\beta)-\zeta(\beta, N+1)}
\end{equation}
Thus, the expected volumes follow the law:
\begin{equation}\label{zipfm}
\bar V_i=\frac{c}{i^\beta}
\end{equation}
The parameters $c$ and $\beta$ are called the \textit{intercept} and the \textit{coefficient} respectively, and in a log-log plot Eq. \ref{zipfm} is a straight line. $c$ is  the expected volume of the most popular query, while $\beta$ characterizes how quickly the volume of queries declines with increasing rank. 

Notice that the $V_i$'s model integer values (the number of searches), whilst the  $\bar V_i$'s may be not an integer. Clearly, when an observed sample of volumes $v_1, \ldots, v_N$ is available, Eq.~\ref{zipfm} is fitted on it (even if the $\bar V_i$'s are not necessarily integers), as for instance it is in Figure \ref{fig:empirical}. Moreover, when all the volumes are observed without noise, the total volume is trivially computed as ${\cal V}=\sum_{i=1}^N v_i$ and the number $N$ of distinct queries is clearly known. The problem becomes much more complicated when: \textit{(i)}  the  volumes are contaminated by some observation noise which biases their observed values, which we model as random variables $X_1, \ldots, X_N$; \textit{(ii)} not all the biased volumes are observed, but only an empirical sample $v_1 \geq v_2 \geq \ldots \geq v_n$ of size $n<N$. In this case the total volume ${\cal V}$ and the number of queries $N$ become random variables: the objective of this paper is to provide an estimation method for ${\cal V}$ and for $N$ assuming a specific distribution of the searches, 
in our case a Zipf's law.
We now present how we model the sources of bias and then introduce the estimation methods.

\begin{figure}[t]
	\centering
	\includegraphics[width=0.40\textwidth]{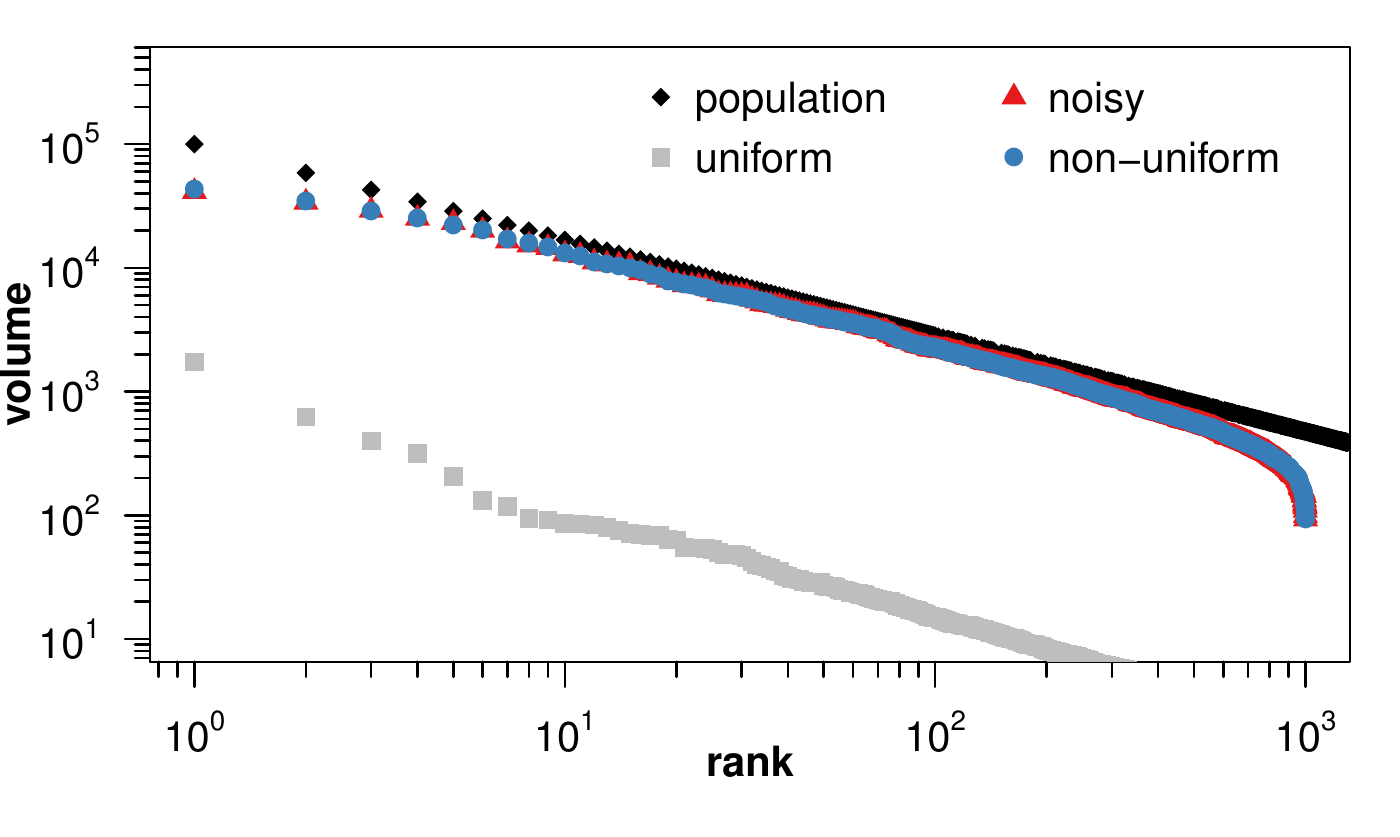}
	\caption{Simulation of different samplings from a Zipf's law.
		\label{fig:simsampling}}
\end{figure}

\begin{figure*}[t]
	\centering
	\includegraphics[trim = 0mm 0mm 0mm 0mm, width=0.40\textwidth]{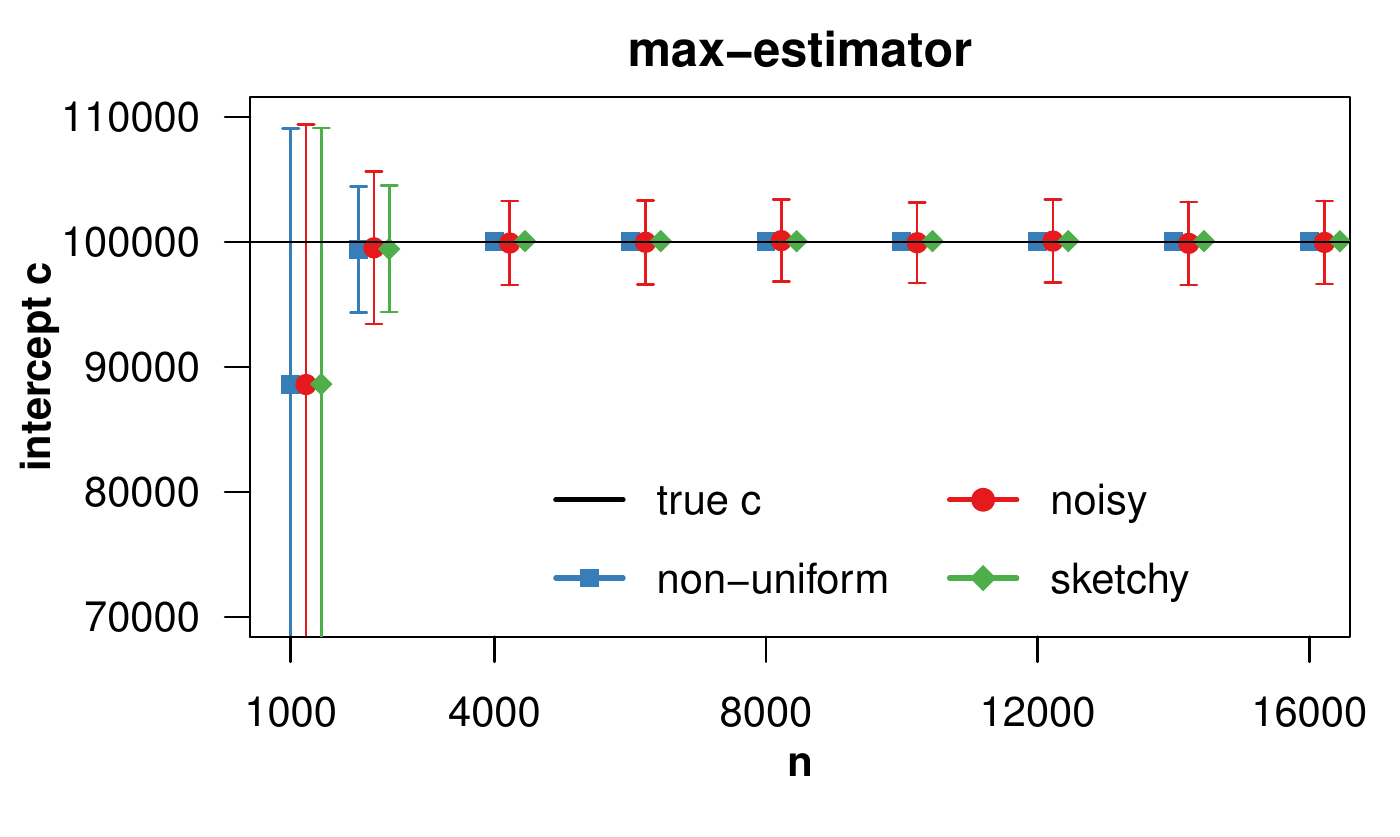}
	\hspace{0.05\textwidth}
    \includegraphics[trim = 0mm 0mm 0mm 0mm, width=0.40\textwidth]{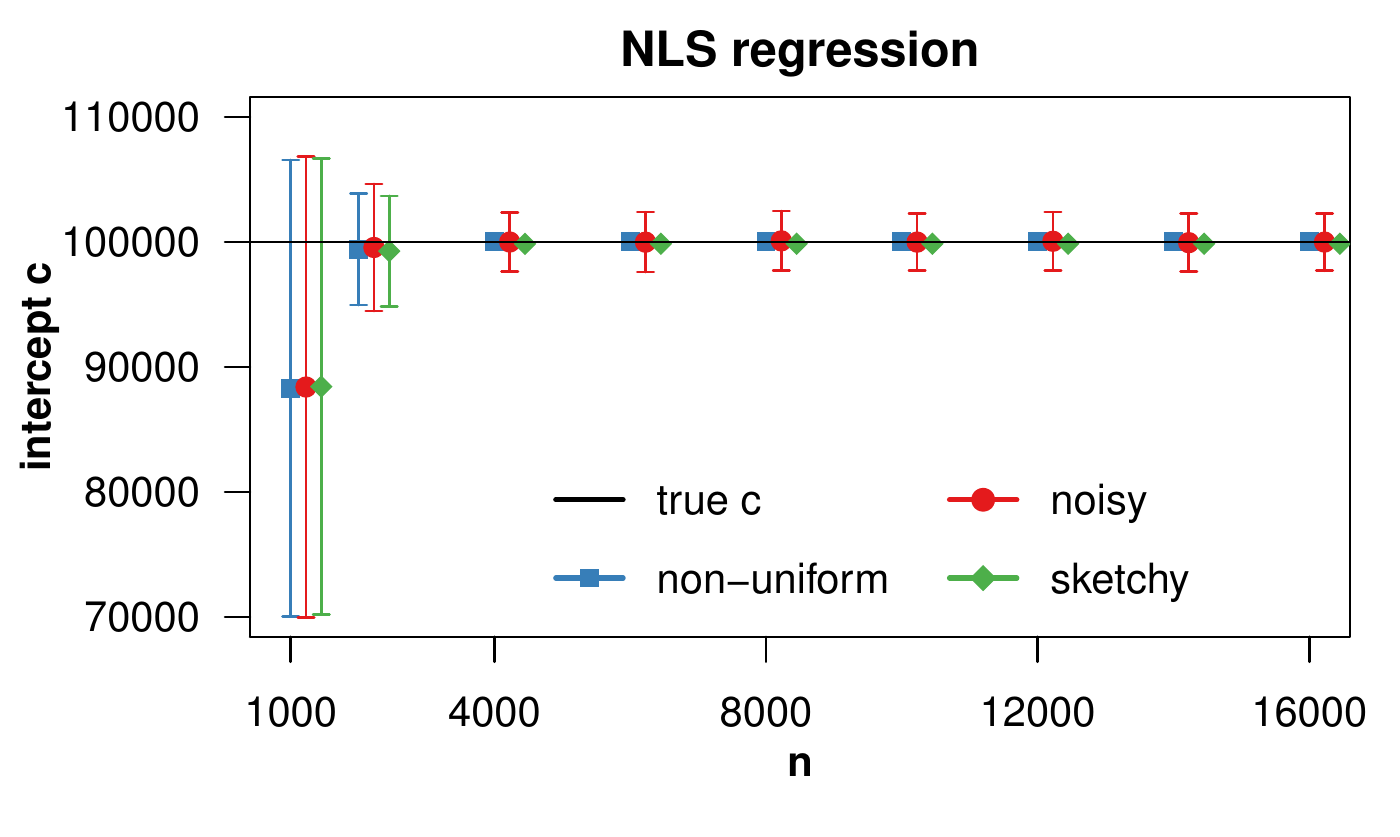}
	\caption{Simulations on  estimation of $c$: error bars (mean $\pm$ stdev). 
		\label{fig:est:c}}
\end{figure*}

\begin{figure*}[t]
	\centering
	\includegraphics[trim = 0mm 0mm 0mm 0mm, width=0.40\textwidth]{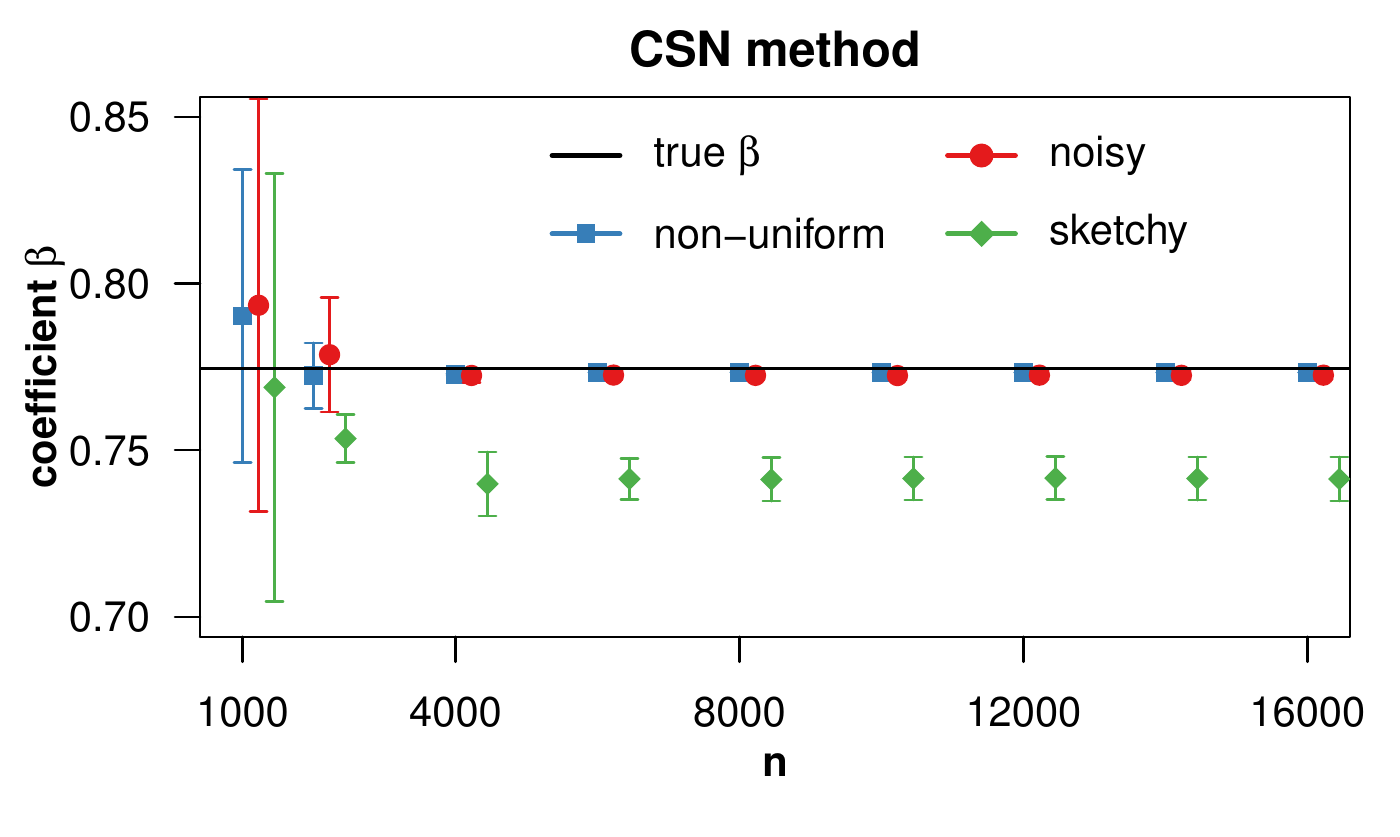}
	\hspace{0.05\textwidth}
	\includegraphics[trim = 0mm 0mm 0mm 0mm, width=0.40\textwidth]{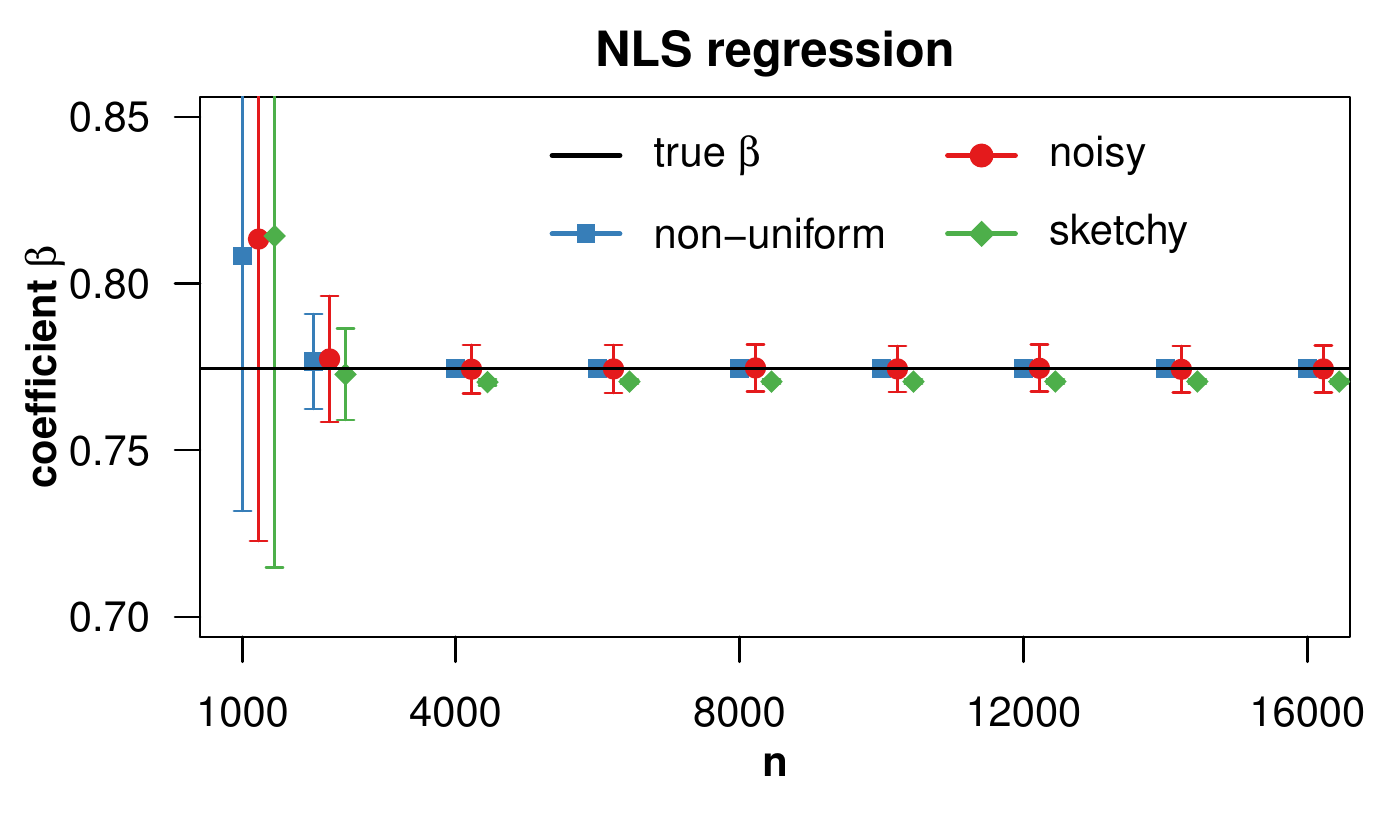}
	\caption{Simulations on  estimation of $\beta$: error bars (mean $\pm$ stdev).
		\label{fig:est:beta}}
\end{figure*}

\section{Biases in sampling from a  Zipf}\label{sec:sampling}

Starting from the assumption that the frequency of searches follows a Zipf distribution (see Eq. \ref{zipf2}), we point out that the empirical distribution in Figure~\ref{fig:empirical} shows a drop of volume in its tail. We investigate on this. We will consider the effects of different sampling methods from a Zipf's law, as well as of noisy samples, and check whether the conclusions are consistent with our empirical data. 

Clearly, uniform sampling from a Zipf's law cannot explain the drop of volume in the tail of the empirical distribution. In fact, queries in an empirical sample are rarely chosen uniformly. The approach followed in our reference domain, for instance, relies on collecting recipe names from specialized websites. The contents of such websites are typically optimized at targeting high-volume keywords through domain-specific keyword research. As a consequence, our empirical data suffers from an unavoidable selection bias in favor of high-volume queries. A similar bias against very low volume queries is introduced by the SEO tools used to obtain observed volumes of queries in a sample. Low volume queries are monitored with low probability by SEO tools. For such queries, SEO tools return no observed volume.
In summary, our empirical data is likely to be a non-uniform sampling of the query population. We assume here that sampling depends on the rank of queries over the population, and call this \textit{non-uniform sampling}. 
Formally, we assume that the $i$-th most searched query is sampled with a probability $p_i$. We want to check whether the observed rank plot over a sample of the population is different from a Zipf's law. To this end, we consider a geometric sampling $p_i\propto p(1-p)^{i-1}$, i.e.,~the sampling
probability decays exponentially with the rank. For example, if $p=0.01$, the probability that the query with the largest volume in the population is observed is $p$, then the second, third, fourth, etc. query in terms of volume will be observed (i.e. sampled) with probability $0.99p$, $0.99^2p$, $0.99^3p$, etc. 
Figure~\ref{fig:simsampling} shows a numerical simulation with the following parameters of the population:
\begin{equation} \label{true_param}
	N=10^6, \quad c=10^5, \quad \beta=0.7745.
\end{equation}
The choice of $\beta$, in particular, has been driven by the empirical distribution of Figure~\ref{fig:empirical}. 
Samples consist of $n=1000$ queries, and $p=0.001$ is set for the geometric sampling. The black line is the whole population, the blue line is obtained with non-uniform sampling while the grey line is obtained with uniform sampling. The non-uniform sampling is consistent with the tail of the empirical distribution in Figure~\ref{fig:empirical}.

As a second bias worth to be considered, SEO tools typically provide approximated values of the true volume of queries, due to a limited user base and/or to computational heuristics in frequency counting.  Therefore, we assume that empirical data is drawn from noisy observed volumes $X_i$'s of the true volumes $V_i$'s. We assume that:
$$
X_i=V_i \epsilon_i
$$
where $\epsilon_i$ are independent observation noise with common distribution characterized by the same mean $\mu$ and variance $\sigma_i^2$. 
Clearly, the presence of noise scrambles the frequencies, and, a fortiori, the ranks of observed volumes may differ from the ranks of true volumes. Figure~\ref{fig:simsampling} includes also a noisy and non-uniform sample (red line) generated assuming $\epsilon_i$ normally distributed, but truncated to 0 to avoid negative $X_i$'s. Parameters are set as follows: $\mu=1$, i.e.,~noise is unbiased, and $\sigma_i^2 = 0.01/9$, i.e.,~$99.7\%$ of noise is in the range $\pm 3 \sigma = \pm 10\%$ of the true value. Noisy and non-uniform sampling (hereafter just \textit{noisy sampling}) produces an empirical distribution very close to the one of non-uniform sampling and that is also consistent with our empirical data.

Another source of bias is due to the use by SEO tools of computationally approximate counting methods \cite{DBLP:journals/ftdb/CormodeGHJ12}. Count-min sketches \cite{DBLP:journals/jal/CormodeM05}, in particular, are extensively used in stream processing. They can be modeled by setting:
\begin{equation}\label{eq:sk}
X_i=V_i + \gamma_i c
\end{equation}
where $\gamma_i$ is uniformly distributed in the range $[0, \gamma]$. In such case, the noise overestimates $V_i$ up to a fraction $\gamma$ of the top volume $V_1 = c/1^\beta = c$. For low volumes, the noise may considerably increase the observed value. However, for a sufficiently low $\gamma$, the non-uniform sampling alleviates from this problem, since low volumes are sampled with low probability. We set $\gamma = 0.001$ in simulations. The empirical distribution generated  lies in between the ones of non-uniform and noisy sampling. For readability reasons, it is not shown in  Figure~\ref{fig:simsampling}. 
We call such model the 
sketchy and non-uniform sampling, hereafter just \textit{sketchy sampling}. %

\begin{figure*}[t]
	\centering
	\includegraphics[trim = 0mm 0mm 0mm 0mm, width=0.40\textwidth]{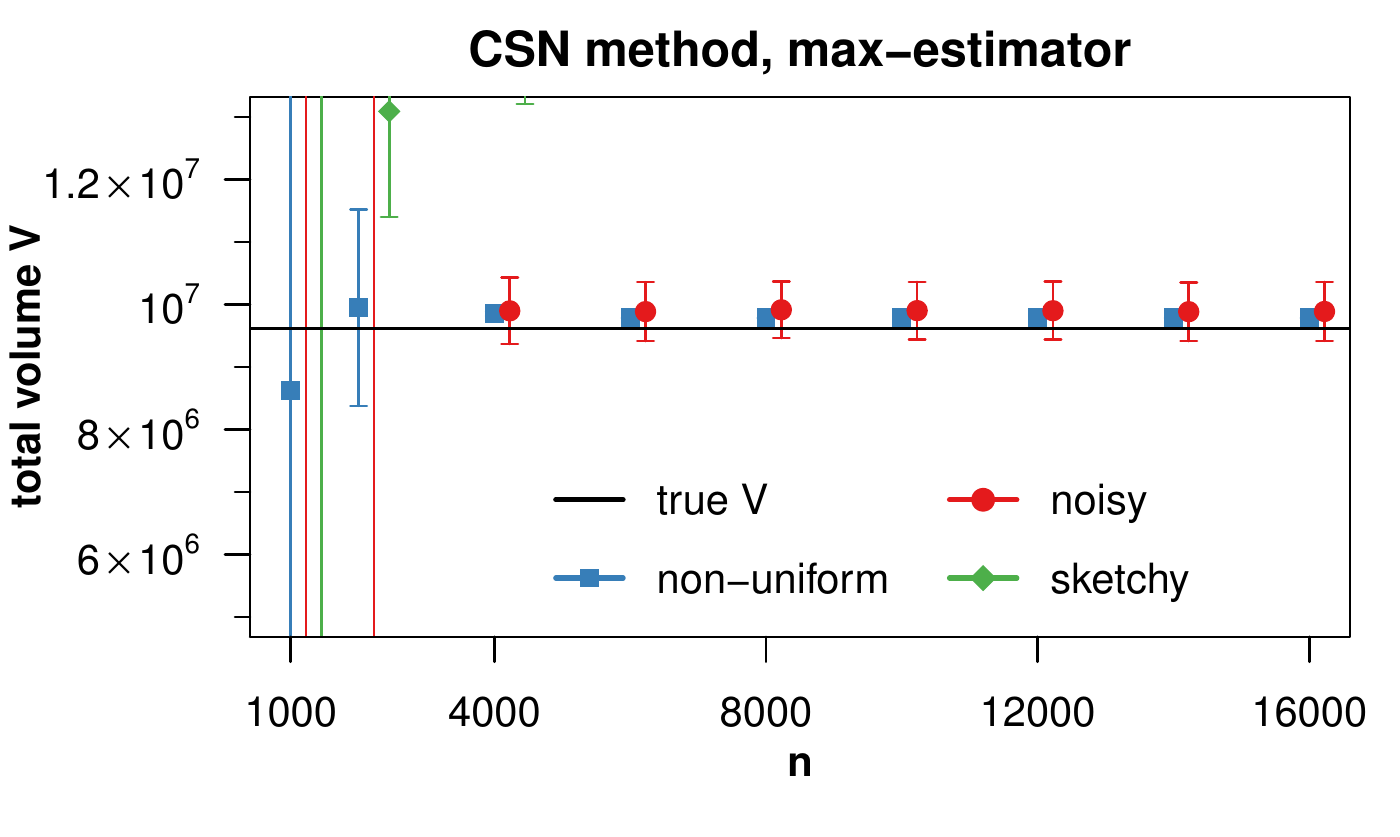}
	\hspace{0.05\textwidth}
	\includegraphics[trim = 0mm 0mm 0mm 0mm, width=0.40\textwidth]{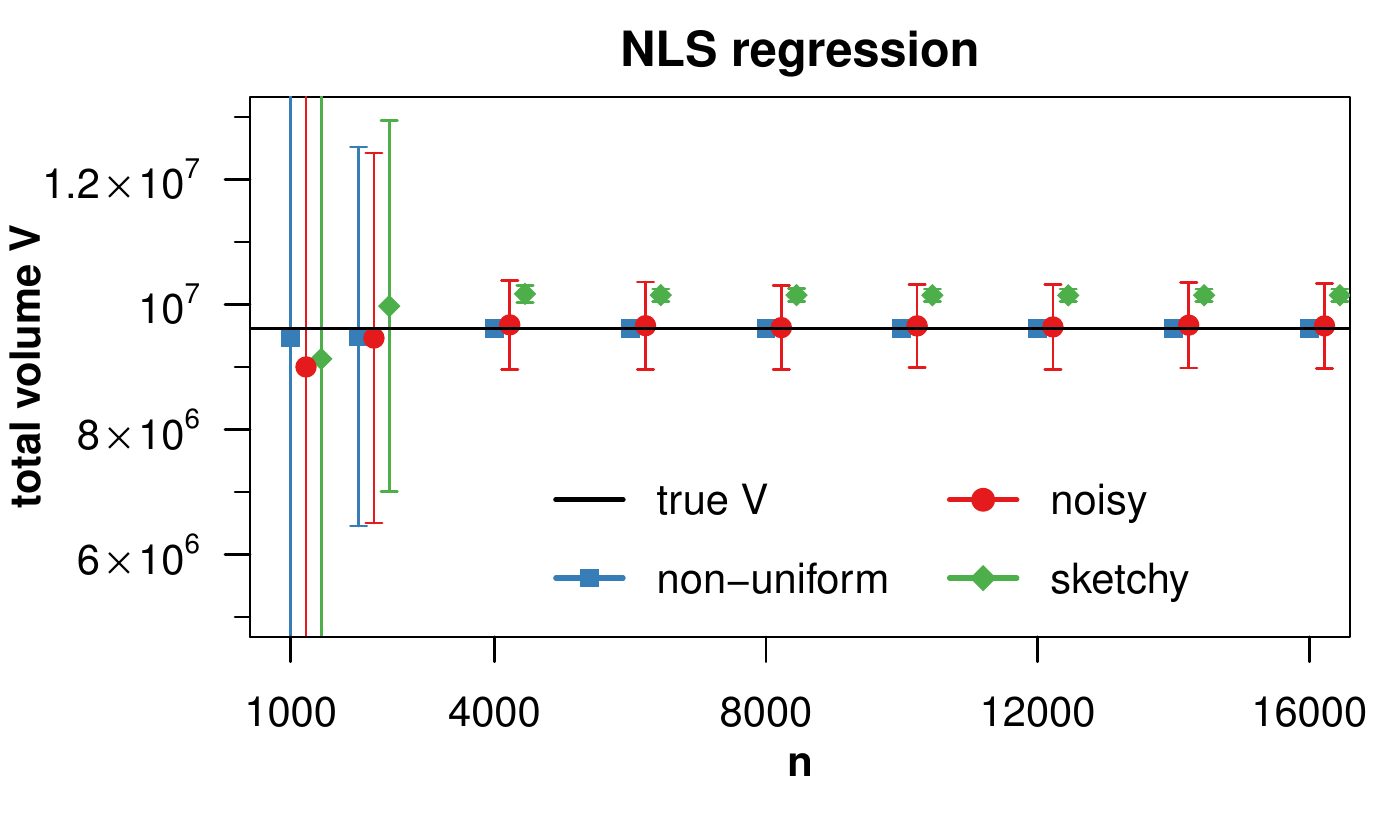}
	\caption{Simulations on the estimation of the total volume $\mathcal{V}$: error bars (mean $\pm$ stdev).  
		\label{fig:est:v}}
\end{figure*}

\section{Estimation methods}\label{sec:model}

\subsection{Estimating $\beta$ and $c$}\label{sec:estbeta}

We start considering the estimation of the coefficient $\beta$ and intercept $c$ in Eq. \ref{zipfm} by exploring two alternative methods for continuous empirical data.
Regarding the coefficient, we observe that $\beta$ is the exponent of the distribution of Eq.~\ref{zipf2}.
Thus, we can rely on the well-known method of Clauset, Shalizi and Newman \cite{DBLP:journals/siamrev/ClausetSN09} (hereafter, the CSN method) for estimating  the $\beta$ parameter in Eq. \ref{zipfm}. 
Strictly speaking, \cite{DBLP:journals/siamrev/ClausetSN09} is a maximum-likelihood estimator $\hat{\alpha}$ of the $\alpha$ exponent of the Power law of volume distribution, $P(V=v) \propto 1/v^{\alpha}$, from the high-volume tail of observed volumes $v_{\mathit{max}} \leq \ldots \leq v_1$. Since in many empirical data the Power law tail is observed only for a range of values, \cite{DBLP:journals/siamrev/ClausetSN09} uses a Kolmogorv-Smirnov like test to determine $v_{\mathit{max}}$ which is the optimal value after which the distribution is Power law tailed. Using the well-known relation $\beta = 1/(\alpha-1)$ between exponents of Power law and continuous Zipf's law (see \cite{DBLP:journals/jasis/Bookstein90,Adamic02}), 
we obtain the estimate $\hat{\beta}= 1/(\hat{\alpha}-1)$ of the coefficient of the rank-volume distribution for top ranks $1$ to $\mathit{max}$. The theoretical advantage of this method is that it automatically selects the rank $\mathit{max}$ from which to regress the coefficient. 

\begin{figure*}[t]
	\centering
	\includegraphics[trim = 0mm 0mm 0mm 0mm, width=0.40\textwidth]{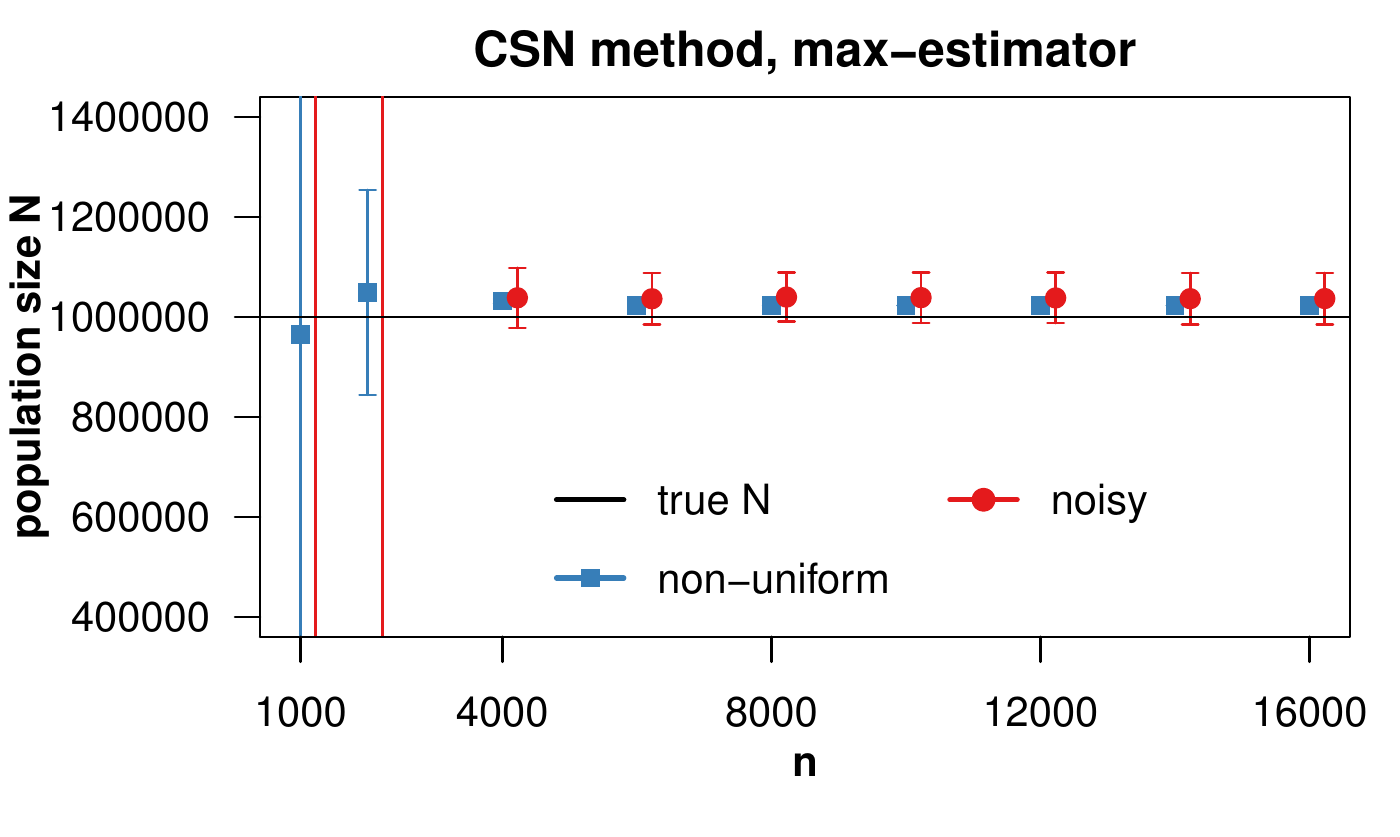}
	\hspace{0.05\textwidth}
	\includegraphics[trim = 0mm 0mm 0mm 0mm, width=0.40\textwidth]{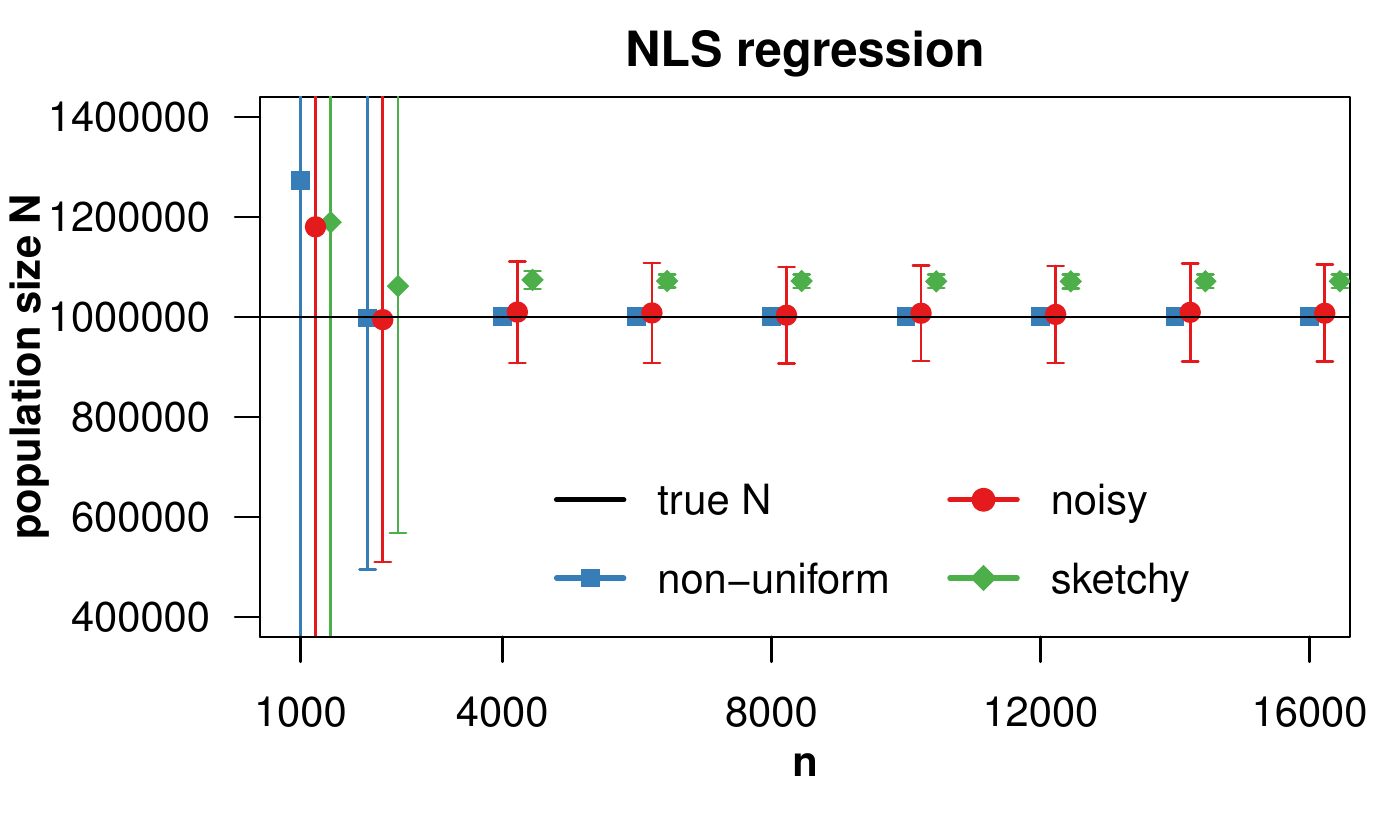}
	\caption{Simulations on  estimation of the population size $N$: error bars (mean $\pm$ stdev). Sketchy case in the left plot exceeds the y-axis limits.
		\label{fig:est:n}}
\end{figure*}

As the second estimator of $\beta$, we use a variant of the standard Nonlinear Least Square (NLS) regression 
of the empirical volume $v_i$ from their rank $i$. The parameters $c$ and $\beta$ are those minimizing the sum of squares:
$\sum_{i=1}^M \left(v_i-\frac{c}{i^{\beta}}\right)^2$, 
where $M$ is the maximal rank considered in the regression\footnote{NLS regression requires to specify initial values for $\beta$ and $c$ to start with. We compute them using OLS regression of the log, i.e. minimizing $\sum (\log v_i-\log c-\beta\log i)^2$. 
OLS regression performs worse than NLS (simulations not shown, see also \cite{epjb}) since it gives too much importance to deviations of low rank with respect to high rank queries.}. Since the empirical data follows a Zipf's law only for the top ranks, we regress the top $M=\mathit{max}$ rank-volume data, where $\mathit{max}$ is the rank returned by the CSN method. In this sense, NLS is a variant of standard Nonlinear Least Square. NLS has two advantages over CSN. First, intercept $c$ and coefficient $\beta$ are estimated together in the same procedure. Second, the regression directly  estimates $\beta$, while in the CSN method $\beta$ is estimated with a formula involving the estimator of $\alpha$. 
Finally, the second estimator of the intercept that we consider here is the maximum observed volume, namely $v_1$. We call it the \textit{max-estimator} of $c$. This is motivated by observing 
that, from Eq.~\ref{zipfm}, $c$ is  the expected volume of the most popular query.

Let us now investigate how those estimators are affected by the non-uniform, noisy, and sketchy sampling from a Zipf's law.
Numerical simulations with parameters as in (\ref{true_param}), are repeated at the variation of the sample size $n$ for $1000$ times and results averaged.

Figure~\ref{fig:est:c} shows that both the max-estimator and the NLS regression converge to the true value of the intercept $c$. For noisy data, however, there is some error, which is proportional to the noise level (set to $\pm 10$\%). Variability is slightly lower for NLS regression. Larger error bars can be observed for small values of $n$. They are due to the chances of not having the queries of the population with the largest volumes included in the sample. 
In Section~\ref{sec:estv}, we will discuss this issue with regard to the estimation of the total volume.
In practical settings, the selection of the sample queries must carefully consider the issue of including in the empirical sample the most popular queries from the population. This has been one of our main concerns in collecting queries in the recipe and cooking domain.

Figure~\ref{fig:est:beta} shows some differences in the estimation of $\beta$. Regarding the CSN method, the estimated values for non-uniform and noisy samplings are slighly lower than the true $\beta$. Underestimation in the sketchy sampling case is, instead, considerable. Regarding the NLS regression, it is unbiased for non-uniform and noisy sampling. For sketchy sampling, $\beta$ is slightly  underestimated. Estimations rapidly converge for increasing $n$'s, except for noisy sampling in the case of NLS, and for sketchy sampling in the case of CSN. 

Finally, all estimations are weakly dependent on $n$:~starting from samples of  $0.4\%$ of the population, they become stable.

\subsection{Estimating $N$} \label{sec:estn}

In the following, we focus on a simple but effective estimator of the size $N$ of the query population. 
We assume that $V_N$, the smallest volume of a query in the population, is known. This assumption is realistic for absolute frequencies, since $V_N = 1$. 
From Eq. \ref{zipfm}, for $i = N$, we have $N = (c/V_N)^{1/\beta}$. This motivates the following estimator:
\begin{equation}
	\label{eq:N}
	\hat{N} = \left(\frac{\hat{c}}{V_N}\right)^{1/\hat{\beta}}
\end{equation}
where $\hat{c}$ is an estimator of $c$, and $\hat{\beta}$ is an estimator of $\beta$. Eq.~\ref{eq:N} can be extended to an estimator of the number of queries whose volume is greater or equal than a given value $v$ as:
\begin{equation}
	\label{eq:Nv}
	\hat{N}_v = (\hat{c}/v)^{1/\hat{\beta}}
\end{equation}

Numerical simulations with parameters as in (\ref{true_param}) are shown in Figure~\ref{fig:est:n} for: (1) $\hat{\beta}$ obtained by the CSN method and $\hat{c}$ obtained by the max-estimator; and (2) $\hat{\beta}$ and $\hat{c}$ obtained by NLS regression. The first method is biased, showing a slight overestimation for non-uniform and noisy sampling and a large overestimation for sketchy  sampling (not shown because exceeding the y-axis limits). The second method converges to the true value of $N$ for non-uniform and noisy sampling (on average), and it slightly overestimates it for sketchy sampling. 
Thus the only advantage of the first method over the second one, is a smaller variability of the estimates in the case of noisy sampling. 

\subsection{Estimating $\mathcal{V}$}\label{sec:estv}

\begin{figure}[t]
		\centering
		\includegraphics[trim = 0mm 0mm 0mm 0mm, width=\columnwidth]{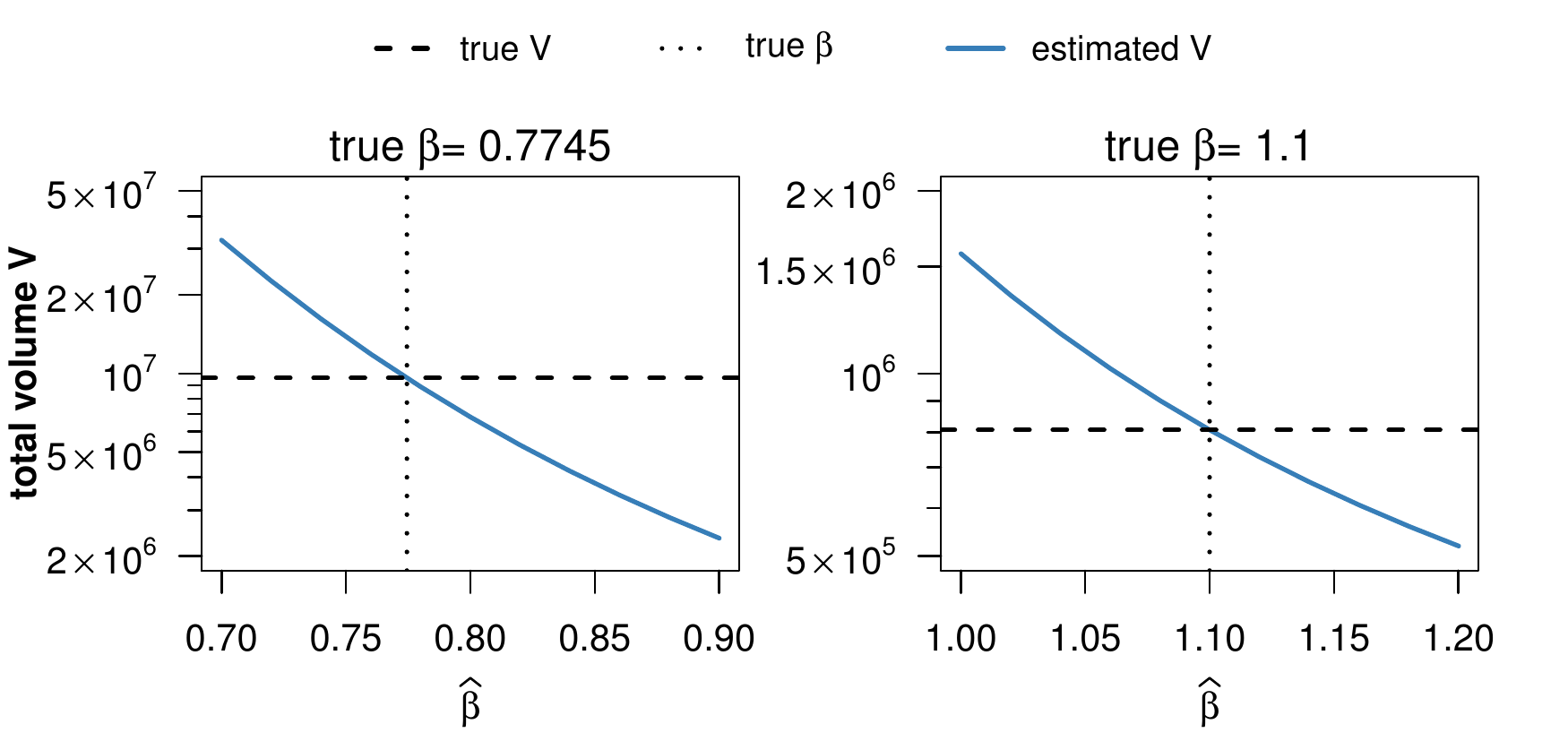}
		\caption{Estimated volume (Eq.~\ref{est:v}) as a function of $\hat{\beta}$, assuming $\hat{c} = c$. Simulation parameters: $N=10^6$, $c=10^5$, $\beta$ in title.  \label{fig:funv}}
\end{figure}

Building on the estimators and simulations conducted so far, the proposed procedure for estimating the total volume $\mathcal{V}$ is composed of the following steps:
\begin{itemize}
	\item estimate $\beta$ and $c$, as described in Section \ref{sec:estbeta};
	\item use the estimated $\hat{\beta}$ and $\hat{c}$  as inputs for estimating $N$ as shown in Section \ref{sec:estn};
	\item the estimator of $\mathcal{V}$ is obtained from Eq. \ref{eq:area} as follows:
	\[ \hat{{\mathcal V}} = \hat{c} [\zeta(\hat \beta)-\zeta(\hat \beta, \hat{N}+1)] \]
\end{itemize}
Notice that by Eq.~\ref{eq:N}, the estimator $\hat{{\mathcal V}}$ can be stated using only $\hat{\beta}$ and $\hat{c}$:
\begin{equation}\label{est:v}
	\hat{{\mathcal V}} = \hat{c} [\zeta(\hat \beta)-\zeta(\hat \beta, \left(\frac{\hat{c}}{V_N}\right)^{1/\hat{\beta}}+1)]
\end{equation}
These estimators can be generalized to estimators of the total volumes of queries with minimum volume $v$ by replacing $V_N$ by $v$:
\begin{equation}\label{est:vv}
	\hat{{\mathcal V}}_v = \hat{c} [\zeta(\hat \beta)-\zeta(\hat \beta, (\hat{c}/v)^{1/\hat{\beta}}+1)]
\end{equation}
Let us continue the previous numerical simulations. With the settings in (\ref{true_param}), it turns out ${\mathcal V} = 9,609,224$. 
First consider using the NLS regression method in the first step of the procedure.
Figure~\ref{fig:est:v} (right) shows that $\hat{\mathcal V}$ converges to ${\mathcal V}$ for non-uniform and noisy sampling, and overestimates it for sketchy sampling. For noisy sampling, there is some variability, which is in the order of the noise introduced during sampling ($\pm 10$\%). The overestimation in the case of sketchy sampling follows from the overestimation of $N$ (see Figure~\ref{fig:est:n}). 

\begin{figure}[t]
	\centering
	\includegraphics[trim = 0mm 0mm 0mm 0mm, width=.42\textwidth]{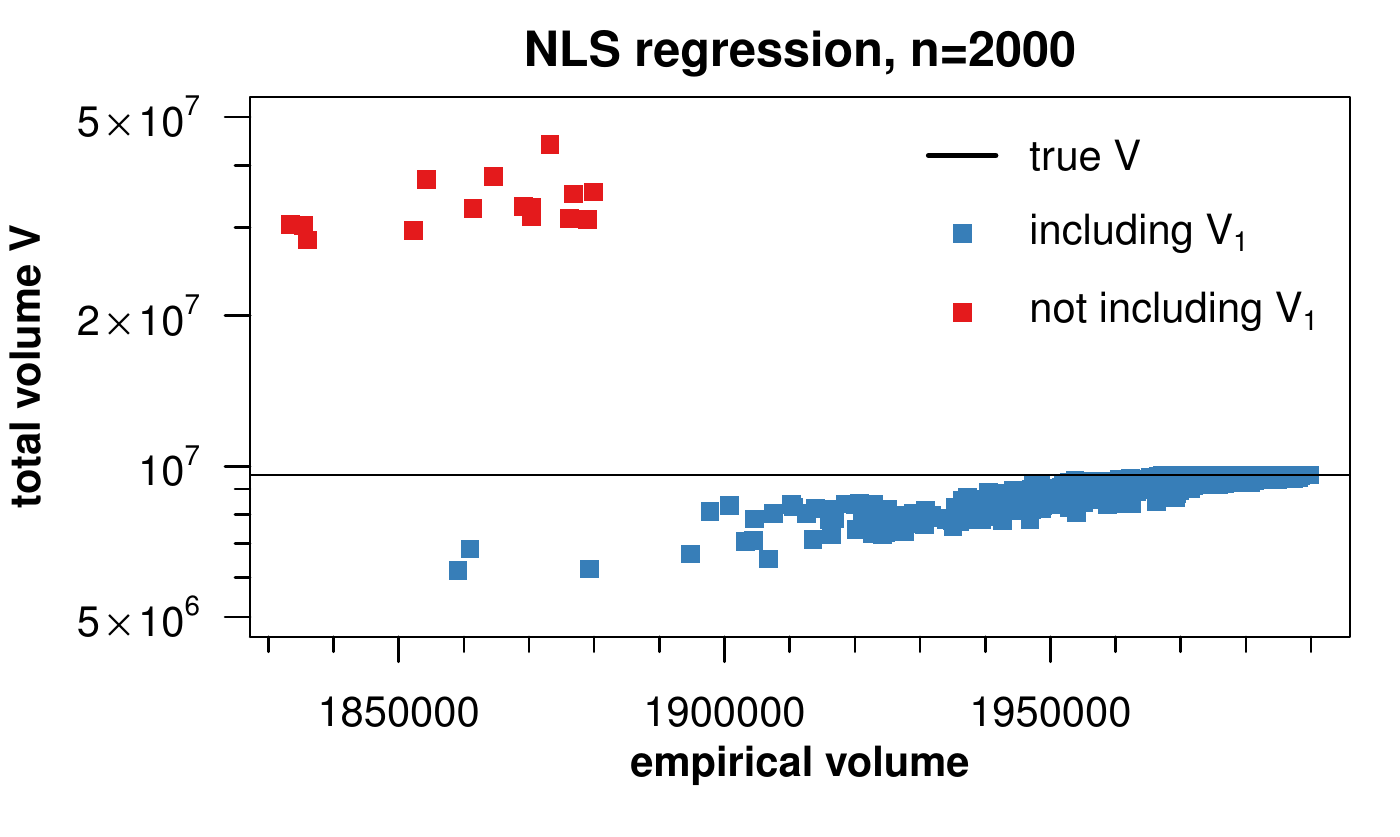}
	\caption{Scatterplot of empirical vs estimated total volume.  \label{fig:volscatter}}
\end{figure}

Consider now the case of using in the first step of the procedure the CSN method coupled with the max-estimator. The total volume shown in Figure~\ref{fig:est:v} (left) is slightly overestimated for non-uniform sampling and for noisy sampling. In the latter case, there is some variability, which appears lower than for the CSN method. This can be tracked back to lower variability in the estimation of $\beta$  (see Figure~\ref{fig:est:beta}). For sketchy sampling, the overestimation is very large: it is out of the bounds of the plot. Again, this can be traced back to a larger underestimation of $\beta$  compared to the CSN method. 

The impact of biased $\hat{\beta}$ on the estimated total volume $\hat{\mathcal V}$ can be readily explained when $\hat{c}=c$ -- which holds in simulations, as shown in Figure~\ref{fig:est:c}. We plot Eq.~\ref{est:v} as a function of $\hat{\mathcal V}$, under the assumption that $V_N$ is known, in Figure~\ref{fig:funv}. The left plot shows simulations for the parameters in (\ref{true_param}) used so far. The right plot uses the same $N$ and $c$, but a $\beta$ greater than $1$. In both cases, the bias of $\hat{\mathcal V}$ is inversely proportional to bias of $\beta$. Note the log scale in the y-axis, which comes from the fact $\beta$ appears as exponent in Eq.~\ref{est:v}. For $\beta$'s lower than $1$, error (or variability) of the estimator $\hat{\beta}$ has a greater impact on error (or variability) of $\hat{\mathcal V}$ than for $\beta$'s greater than $1$.

\begin{figure*}[t]
	\centering
	\includegraphics[trim = 0mm 0mm 0mm 0mm, width=0.40\textwidth]{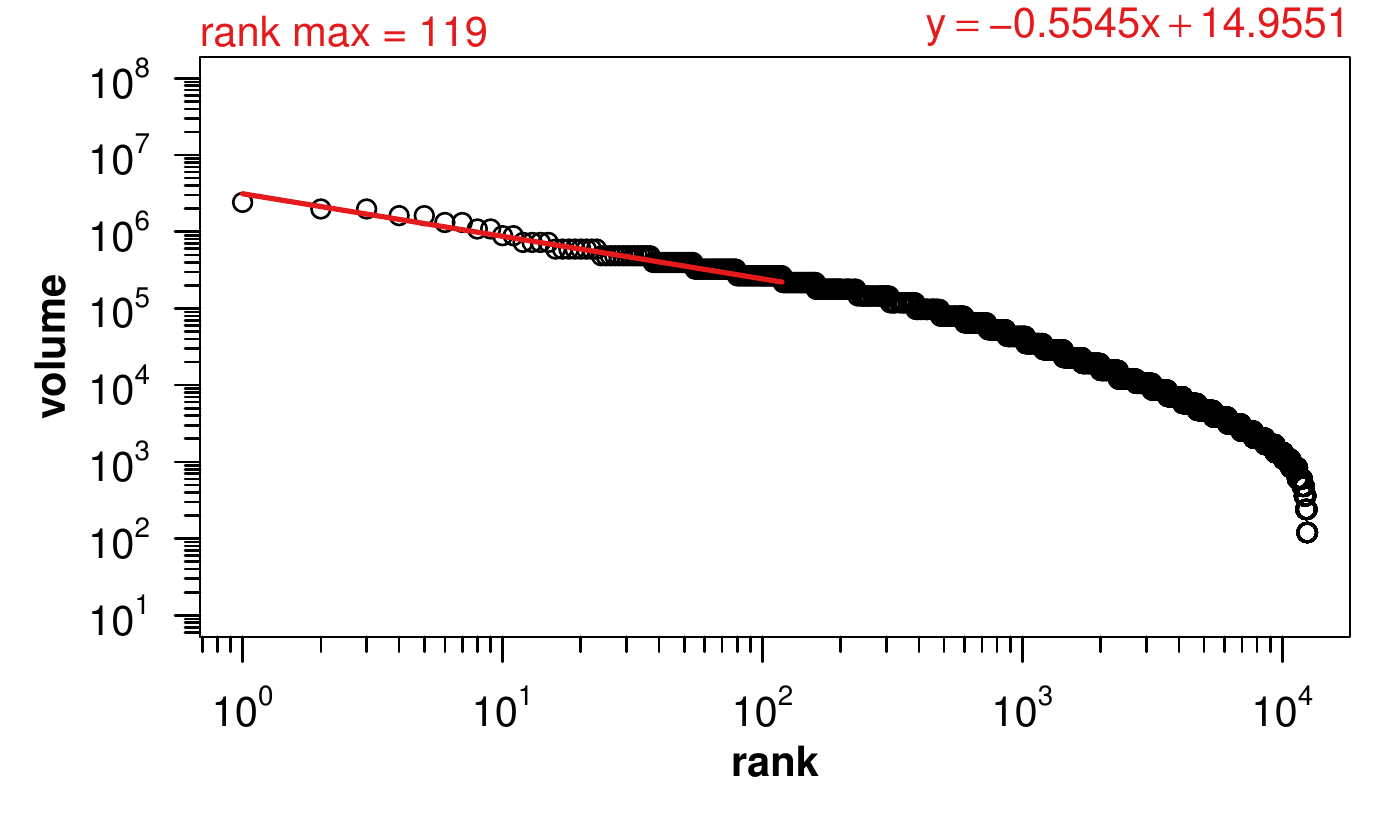}\hspace{0.05\textwidth}
	\includegraphics[trim = 0mm 0mm 0mm 0mm, width=0.40\textwidth]{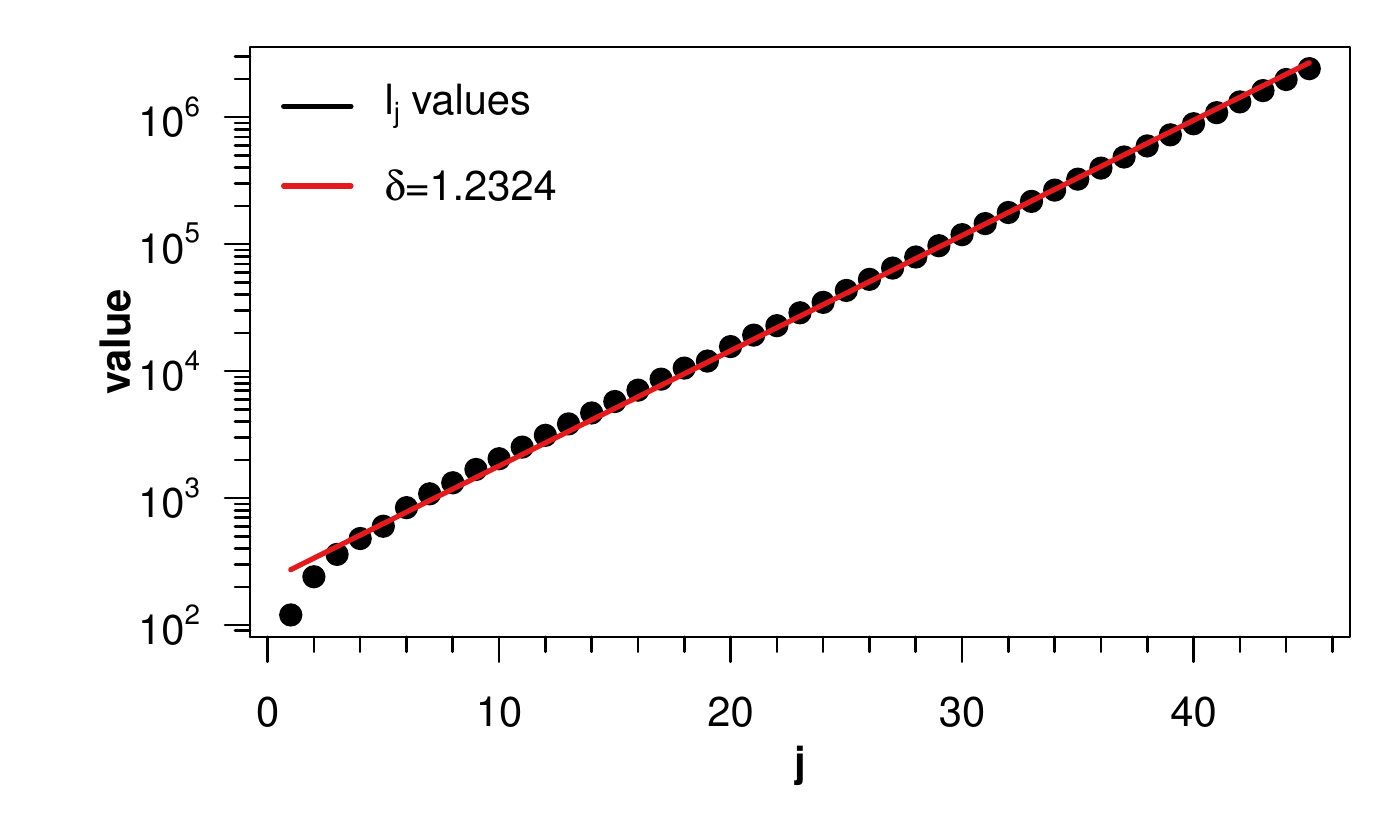}
	\caption{Left: empirical rank-volume distribution (SearchVolume estimates). Right: bins from the empirical distribution.
		\label{fig:semrush}
		\label{fig:bin}}
\end{figure*}

We already observed that the performances of the estimators become stable from $n=4,000$ on, which is $0.4$\% of the population size. Let us now focus on smaller sample sizes, for which instead there is a large standard deviation over the experimental runs. Fix $n=2,000$, and consider NLS regression and non-uniform sampling. 
From Figure~\ref{fig:est:v} (right), we have that the standard deviation of the estimates $\hat{\mathcal V}$ over the 1,000 experimental runs is approximately $3\times10^6$. What is the source of such variability?  Figure~\ref{fig:volscatter} shows the scatter plot of the estimated total volume vs the empirical volume (i.e.,~the sum of observed volumes) of the sample for each of the 1,000 runs. Runs with a lower empirical volume exhibit most of the variability (notice that the y-axis is in log-scale). If the empirical volume is sufficiently large, the estimated total volume converges to the true volume. Moreover, if the sample includes the query with the largest volume in the population $V_1$ (blue points in Figure~\ref{fig:volscatter}), the estimate is less biased than in the case the sample does not include $V_1$ (red points). Not having $V_1$ in the sample causes underestimation of $\beta$, which in turn causes overestimation of the total volume. Overall, this reinforces our previous conclusion that, in practical settings, the selection of sample queries must carefully include the most popular ones.

As a summary of the simulations, we therefore recommend using the NLS regression method for estimating $c$ and $\beta$, and, using Eqs.~\ref{est:v}--\ref{est:vv}, for estimating ${\mathcal V}$ and ${\mathcal V}_v$.

\subsection{Errors on the estimates}\label{sec:errors}
We now compute the error on the estimated $N$ obtained from Eq.~\ref{eq:N}. Using the propagation of errors under the assumption that the errors $\Delta \beta$
on $\beta$ and $\Delta c$ on $c$ are independent, the error on $\hat N$ is:
$$
\Delta N=\sqrt{\left(\frac{\partial \hat N}{\partial \hat c}\Delta c\right)^2+\left(\frac{\partial \hat N}{\partial \hat \beta}\Delta \beta\right)^2}
$$
The values $\Delta c$ and $\Delta \beta$ are set to the standard errors of the parameter estimation. In particular, for NLS regression they are directly provided by the \textit{nls()} function of the R \textit{stats} package, which uses a linearization approach\footnote{\href{http://sia.webpopix.org/nonlinearRegression.html\#standard-errors-of-the-parameter-estimates}{\textit{http://sia.webpopix.org/nonlinearRegression.html\#standard-errors-of-the-parameter-estimates}}}.
To have a more conservative estimate of $\Delta N$, taking into account correlations between errors, one can replace the previous formula with the sum of the absolute values:
\begin{equation}\label{eq:errN}
	\Delta N=\left|\frac{\partial \hat N}{\partial \hat c}\Delta c\right|+\left|\frac{\partial \hat N}{\partial \hat \beta}\Delta \beta\right|
\end{equation}
The partial derivatives in the previous expression are:
\begin{equation}\nonumber
	\frac{\partial \hat N}{\partial \hat c}=\left(\frac{\hat{c}}{V_N}\right)^{1/\hat{\beta}}\frac{1}{\hat \beta \hat c}=\frac{\hat N}{\hat \beta \hat c}
\end{equation}
\begin{equation}\nonumber
	\frac{\partial \hat N}{\partial \hat \beta}=-\left(\frac{\hat{c}}{V_N}\right)^{1/\hat{\beta}}\frac{1}{\hat \beta^2}\log\frac{\hat c}{V_N}= -\frac{\hat N}{\hat \beta^2}\log\frac{\hat c}{V_N}
\end{equation}
Similarly, the error on the total volume is: 
\begin{equation}\label{eq:dV}
	\Delta {\mathcal V}=\left|\frac{\partial \hat {\mathcal V}}{\partial \hat c}\right| \Delta c+\left|\frac{\partial \hat {\mathcal V}}{\partial \hat \beta}\right |\Delta \beta
\end{equation}    
The calculation of the partial derivatives is a bit more involved. We start from Eq.~\ref{est:v}, where $\hat {\mathcal V}$ is a function of $\hat c$ and $\hat \beta$.
We find:
{\small
\begin{equation*}
	\frac{\partial {\mathcal V}}{\partial \hat{c}}=\frac{{\hat{\mathcal V}}}{\hat{c}}+\hat{N}\zeta\left(\hat{\beta}+1,\hat{N}+1\right)
\end{equation*}
\begin{equation*}
	\frac{\partial {\mathcal V}}{\partial \hat{\beta}}=\hat{c}\left(\zeta'(\hat{\beta})-\zeta^{(1,0)}(\hat{\beta},\hat{N}+1)-\frac{\hat{N}\log(\frac{\hat{c}}{V_N})\zeta(\hat{\beta}+1,\hat{N}+1)}{\hat{\beta}}\right)
\end{equation*}
}
where $\zeta'(x)$ is the derivative of the Riemann Zeta function and $\zeta^{(1,0)}(s,a)$ is the partial derivative of the Hurwitz function with respect to $s$. 

\begin{figure}[t]
	\centering
	\includegraphics[width=0.40\textwidth]{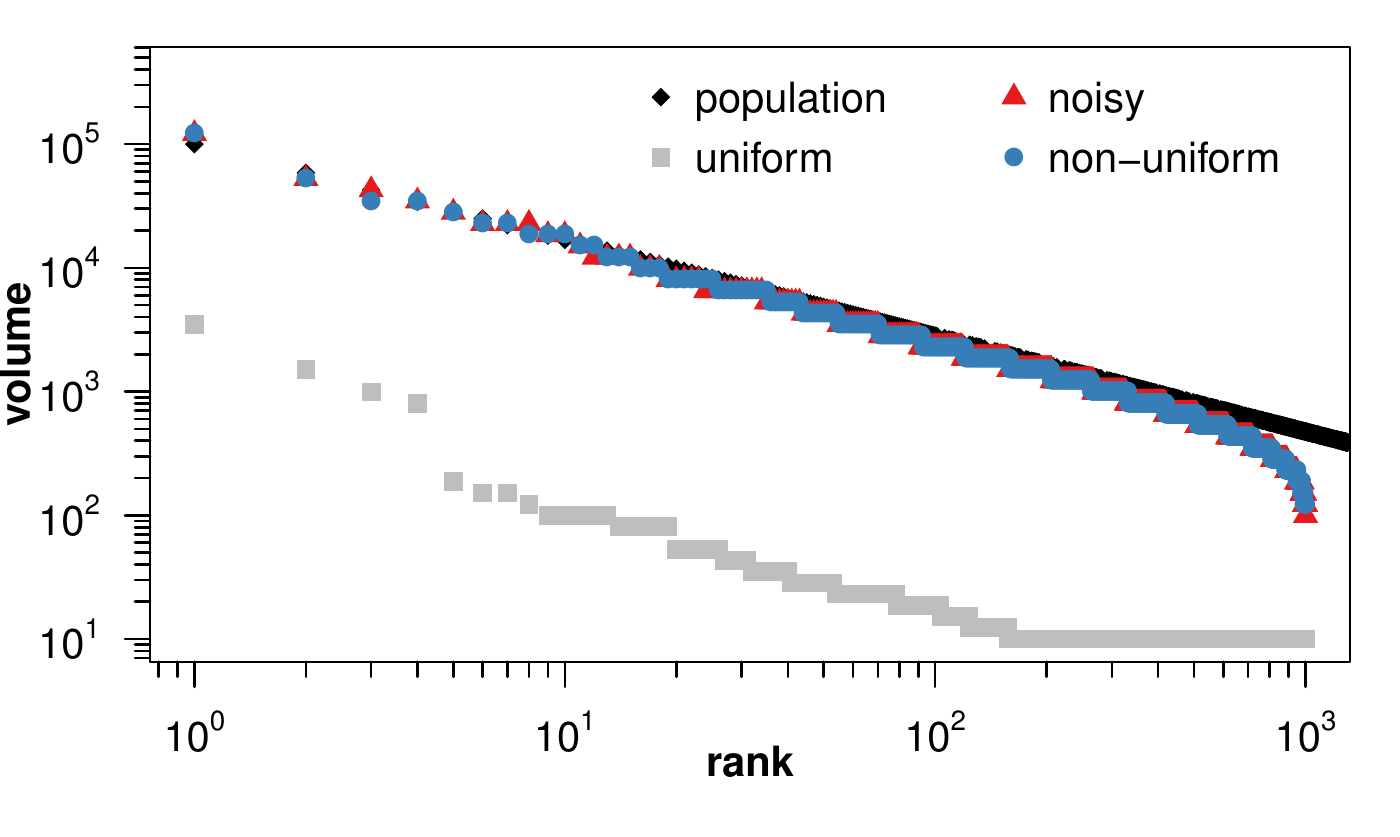}
	\caption{Simulation of binned sampling from a Zipf's law.\label{fig:simsampling:binned}}
\end{figure}

\section{The case of binned data}\label{sec:binned}

Most SEO tools do not provide an absolute observed volume of a query, but rather they provide an interval estimate of the volume, i.e.,~binned observed volumes. The motivation is that intervals ameliorate for the noise introduced in the estimation process. 
Figure~\ref{fig:semrush}~(left) shows the rank-volume distribution of a sub-sample of recipes queries whose binned observed volumes are obtained from the SearchVolume tool. 
Differences with Google Trends distribution will be discussed later on in Section~\ref{sec:emp:sv}. For now, we observe that the rank-volume distribution is still Zipfian.

\begin{figure*}[t]
	\centering
	\includegraphics[trim = 0mm 0mm 0mm 0mm, width=0.40\textwidth]{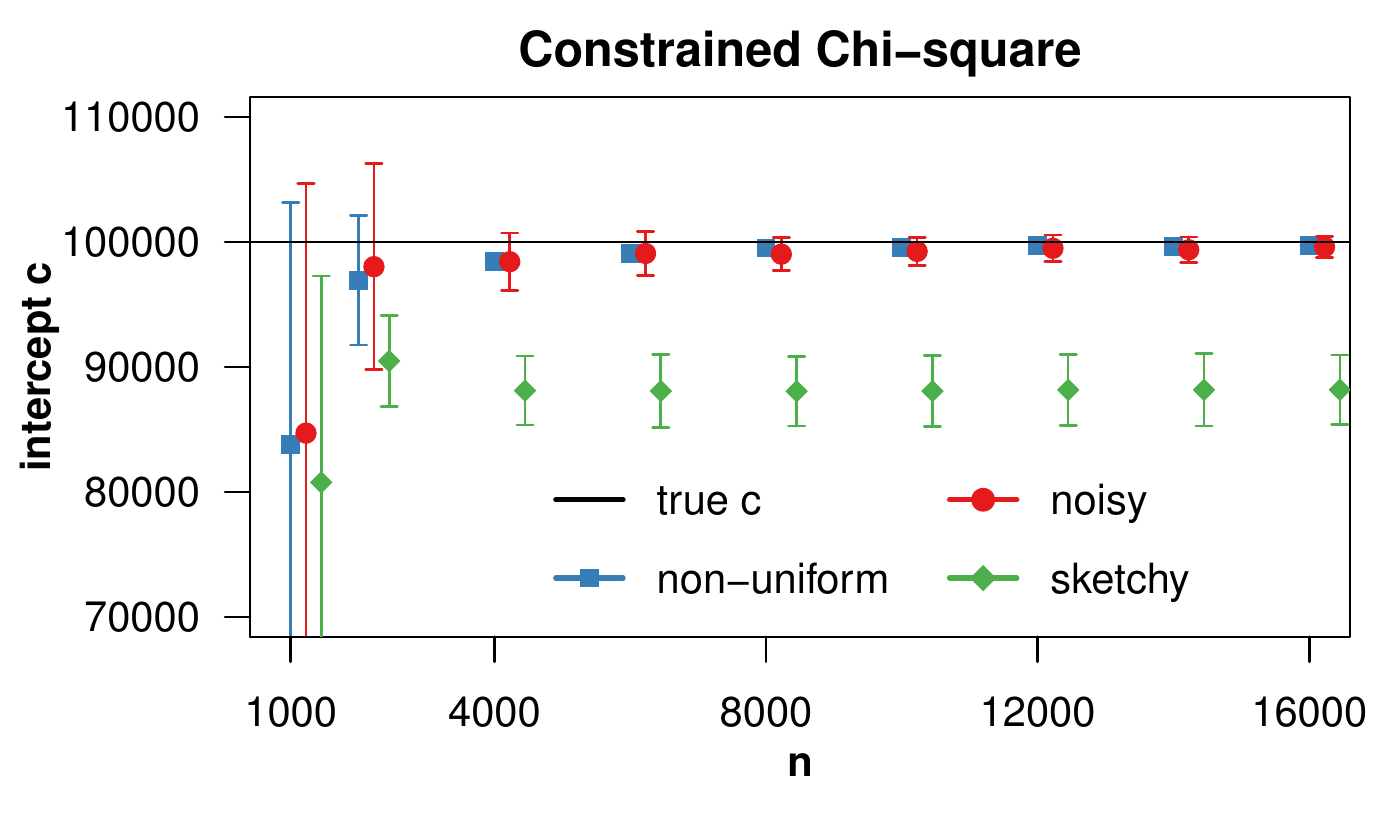}
	\hspace{0.05\textwidth}
	\includegraphics[trim = 0mm 0mm 0mm 0mm, width=0.40\textwidth]{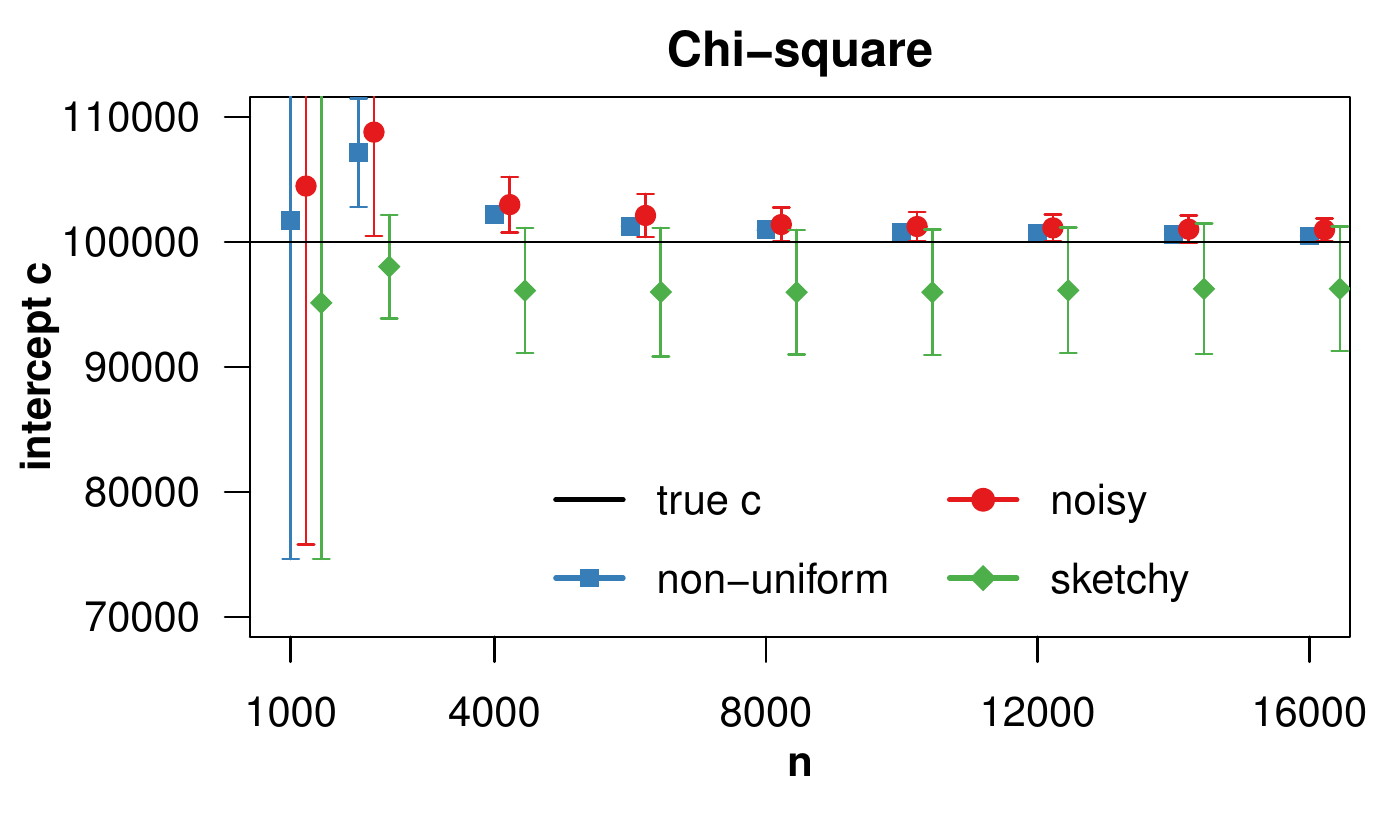}
	\caption{Simulations on estimation of $c$ for binned data: error bars (mean $\pm$ stdev).	The horizontal line is the true value of $c$.\label{fig:est:binned:c}}
\end{figure*}
\begin{figure*}[t]
	\centering
	\includegraphics[trim = 0mm 0mm 0mm 0mm, width=0.40\textwidth]{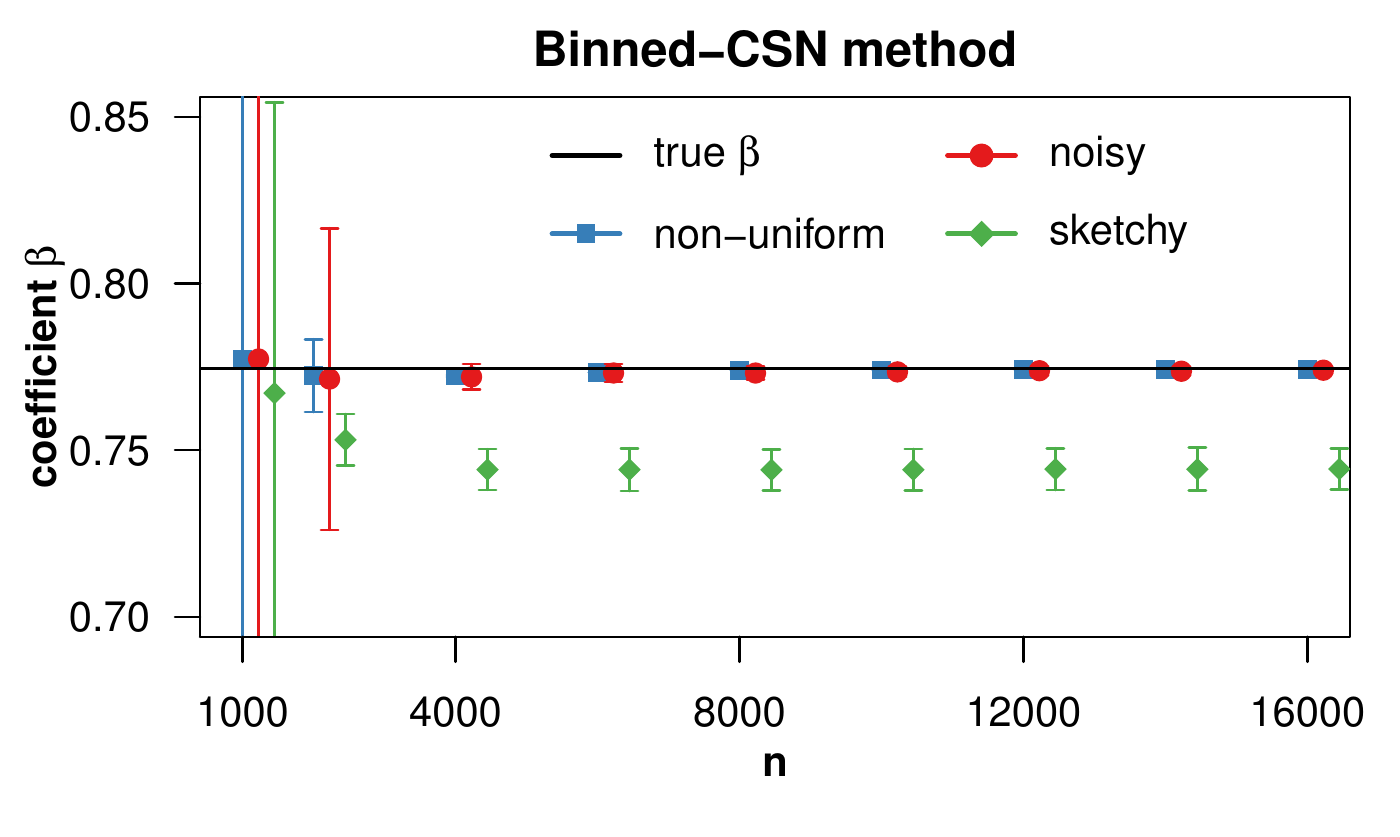}
	\hspace{0.05\textwidth}
	\includegraphics[trim = 0mm 0mm 0mm 0mm, width=0.40\textwidth]{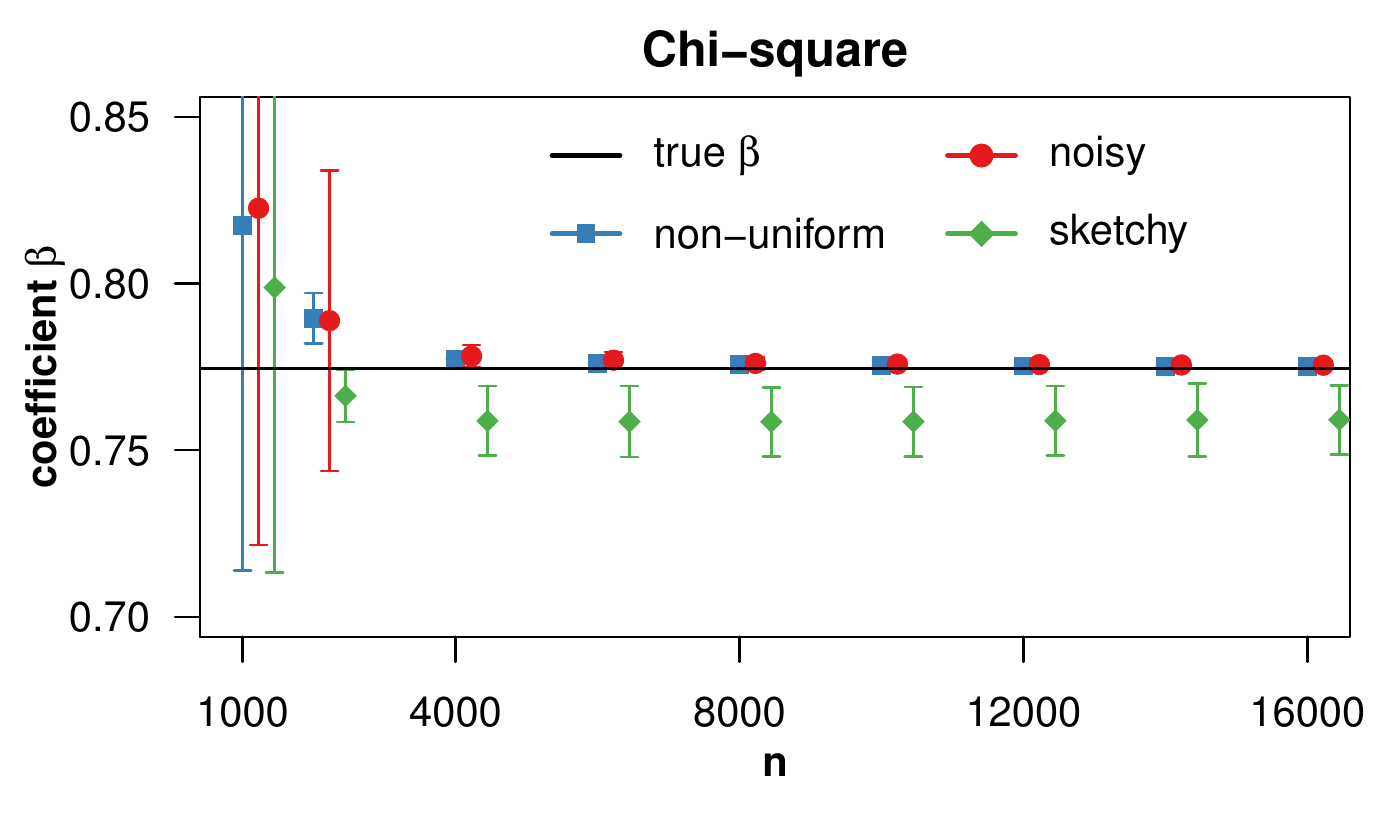}
	\caption{Simulations on estimation of $\beta$ for binned data: error bars (mean $\pm$ stdev). The horizontal line is the true value of $\beta$.	\label{fig:est:binned:beta}}
\end{figure*}

The binned nature of data  demands for specific estimation methods. For example, \cite{VirkarClauset2014} generalizes the CSN method to the estimation of the coefficient of Power law distributed data if these are binned.
We extend here the approach of the previous section to the case of discrete data obtained by binning (possibly noisy or sketchy) values. 
Specifically, we consider two strategies, one based on the method from \cite{VirkarClauset2014} and the other based on Chi-square minimization.

As before, we assume that the frequency of searches follow a Zipf law (see Eq.~\ref{zipf2}), but the observed volumes are binned according to some scheme. 
We assume that there are $M$ bins, and that for $j \in [1, M]$, the $j$-th bin consists of the interval $[\ell_{j-1},\ell_j)$, where $\ell_0 = V_N$ (the smallest volume). SEO tools typically report as observed volume the upper bound $\ell_{j}$ of the bin. For instance, Figure~\ref{fig:bin}~(right), reports the $\ell_j$'s values for the SearchVolume data shown in the left panel of the figure. They approximately follow a geometric progression, namely: 
\begin{equation}\label{eq:geomprogr}
\mbox{\rm for $j \in [1, M]$}\ \ell_{j} = \ell_1 \delta^{j-1}.
\end{equation} 
Under such a binning scheme, 
a continuous volume $V \geq V_N$ belongs to the $j_V$-th bin, where:
$$
j_V = \mathit{max}\left\{1, 2+\left\lfloor\log_\delta \frac{V}{\ell_1}\right\rfloor\right\}
$$
and its binned volume is $\ell_{j_V} = \ell_1 \delta^{\mathit{max}\left\{0, 1+\left\lfloor\log_\delta \frac{V}{\ell_1}\right\rfloor\right\}}$.

In simulations throughout this section, we apply such a discretization scheme to the non-uniform, noisy, and sketchy continuous sample data with parameters as in  (\ref{true_param}), and keep the same nomenclature for the discretized versions of those sampling methods. Figure~\ref{fig:simsampling:binned} shows the impact of sampling, for the same settings of Figure~\ref{fig:simsampling} but with geometric binning where:
\begin{equation}\label{eq:delta}
\delta = 1.2324
\end{equation}
is the ratio used by SearchVolume, as shown in Figure \ref{fig:bin}~(right). As for the continuous case, uniform sampling is not consistent with empirical data, while the other strategies are.

\subsection{Binned-CSN}

Ref. \cite{VirkarClauset2014} extends to the case of binned data the original CSN approach of using a maximum likelihood estimator of the exponent and a Kolmogorov-Smirnov test for selecting the tail of values that best fit a Power law. As a consequence, we can extend our CSN-based estimator $\hat{\beta}= 1/(\hat{\alpha}-1)$ to binned data, where $\hat{\alpha}$ is the estimator of the exponent of a binned Power law. 

As in the continuous case, we will also make use in the next subsections of the tail of the (binned) values $v_{\mathit{max}} \leq \ldots \leq v_1$ that best fit a Power law. This corresponds to consider  bins $[j_{v_{\mathit{max}}}, M]$, where $j_{v_{\mathit{max}}}$ is the bin of $v_{\mathit{max}}$.

Simulation results in Figure~\ref{fig:est:binned:beta}~(left) show that Binned-CSN performs very well for non-uniform and noisy sampling, while it under-estimates $\beta$ for sketchy sampling. 

\begin{figure*}[t]
	\centering
	\includegraphics[trim = 0mm 0mm 0mm 0mm, width=0.40\textwidth]{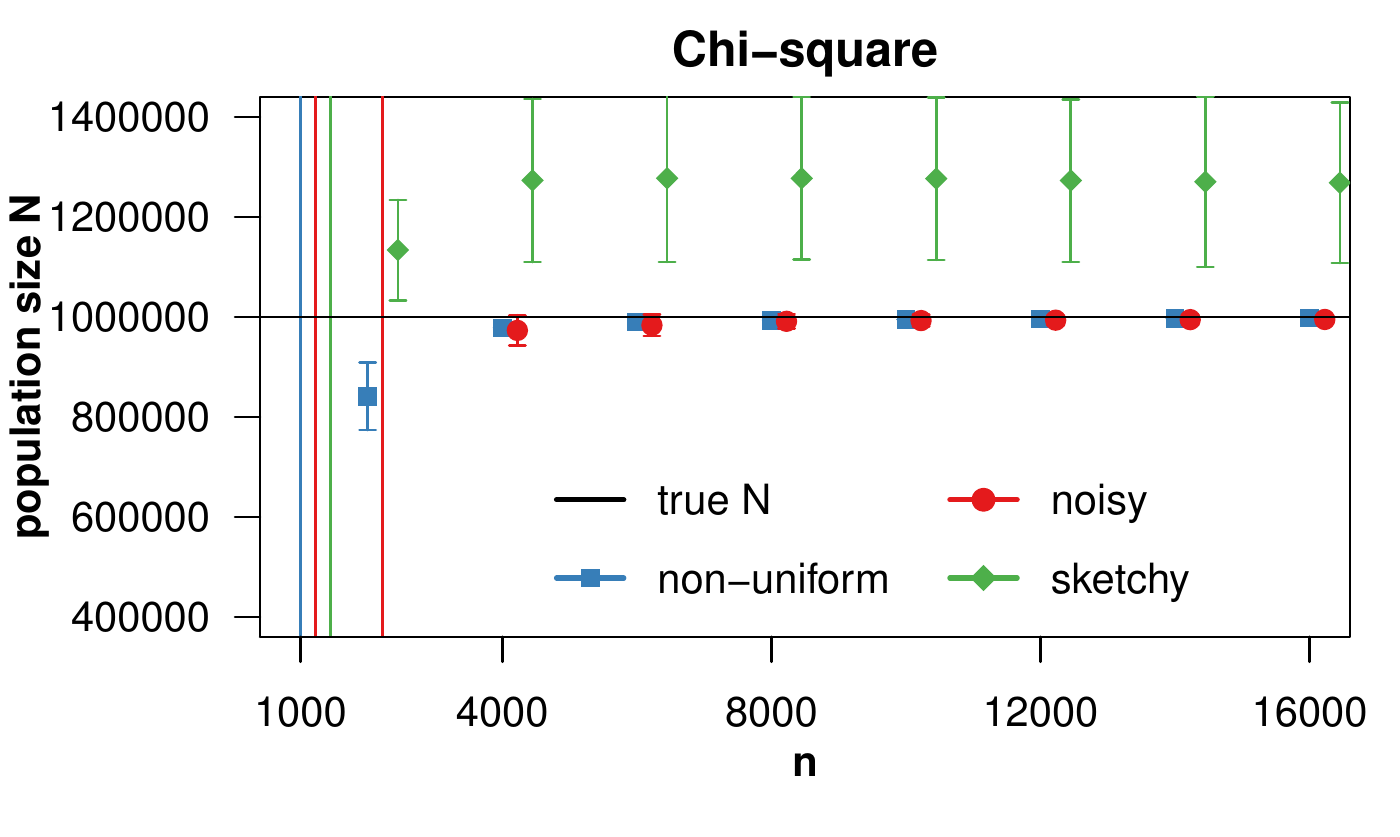}
	\hspace{0.05\textwidth}
	\includegraphics[trim = 0mm 0mm 0mm 0mm, width=0.40\textwidth]{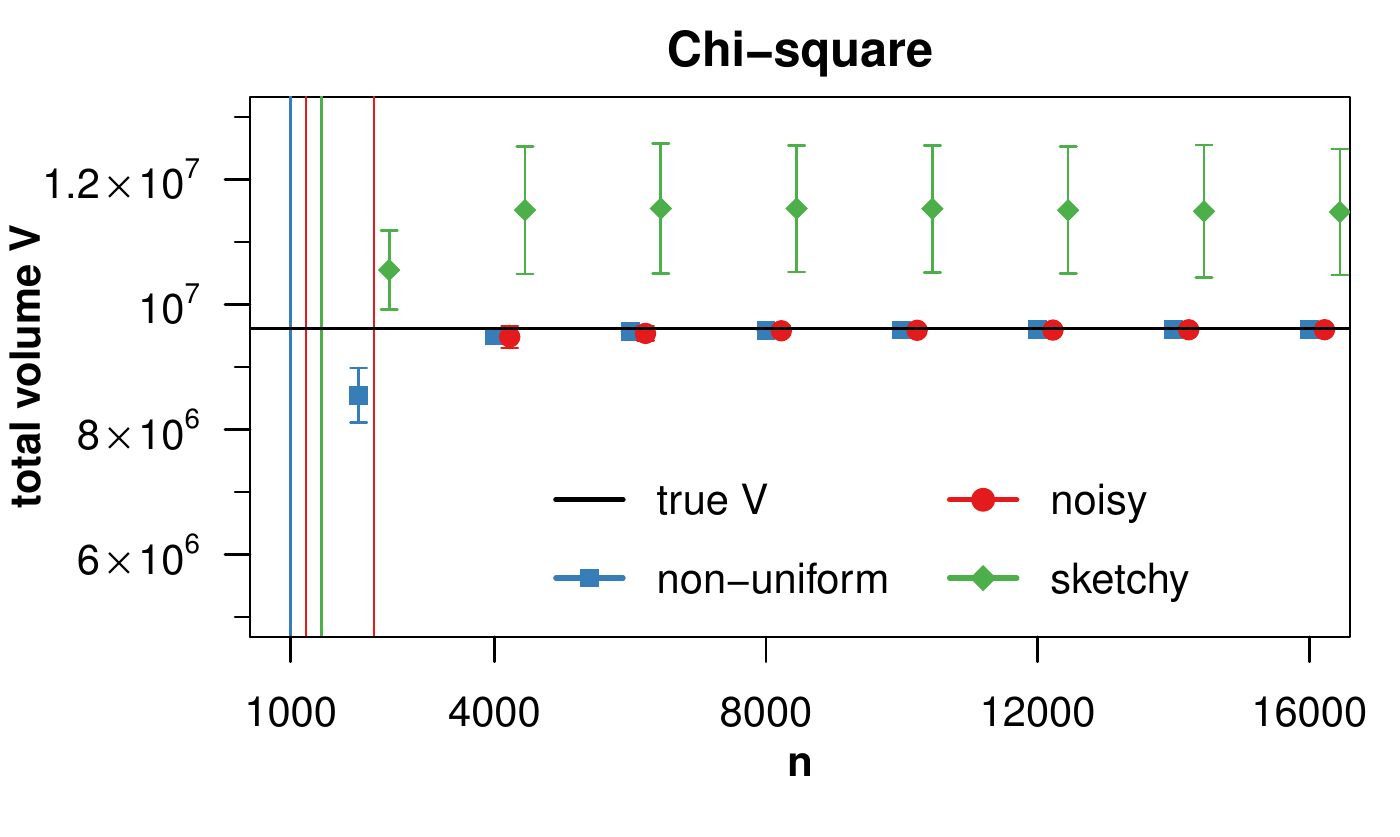}
	\caption{Simulations on estimation of $N$ and ${\cal V}$ for binned data: error bars (mean $\pm$ stdev). The horizontal lines are the true value of $N$ and $\mathcal{V}$.	\label{fig:est:binned:nv}}
\end{figure*}
\begin{figure*}[t]
	\centering
	\includegraphics[trim = 0mm 0mm 0mm 0mm, width=0.40\textwidth]{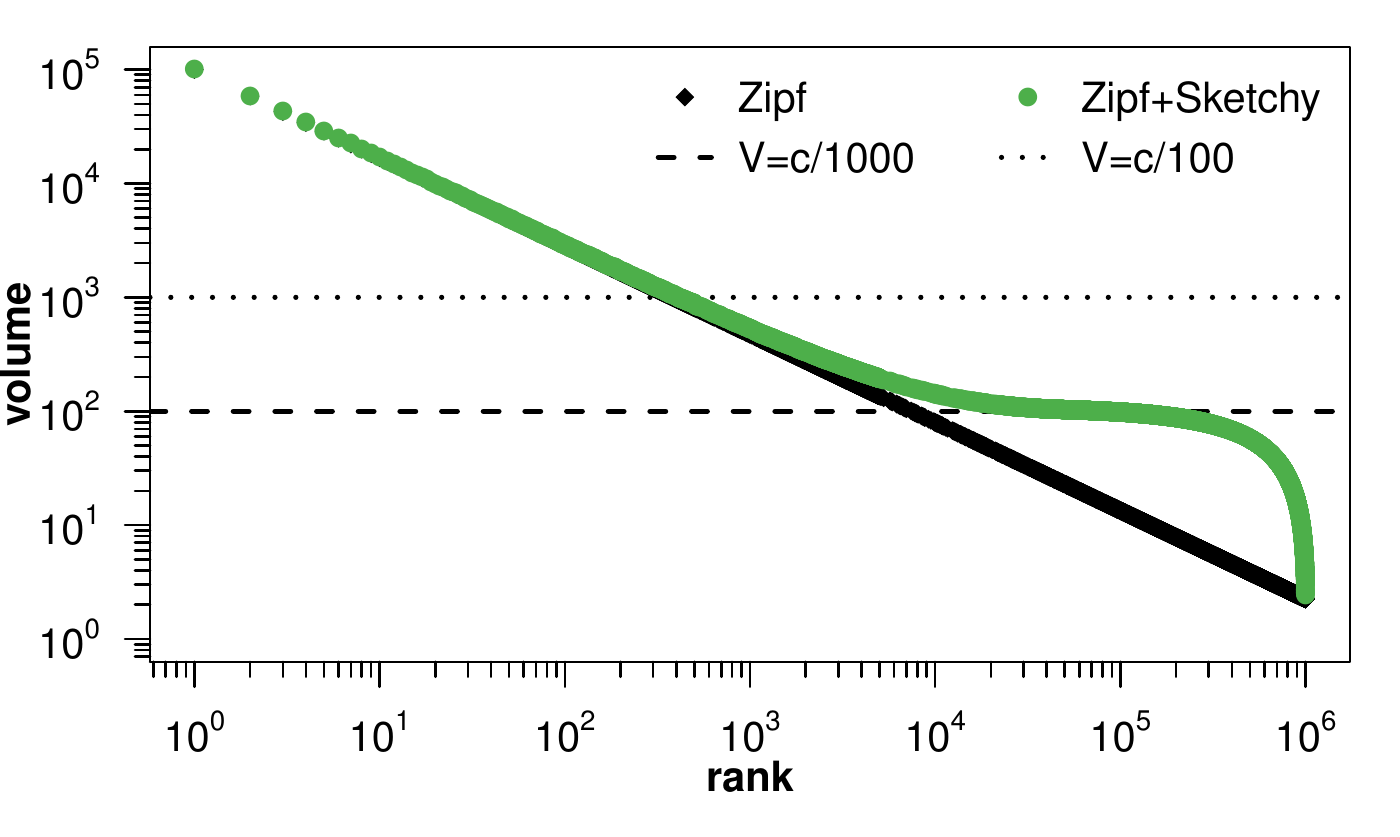}
	\hspace{0.05\textwidth}
	\includegraphics[trim = 0mm 0mm 0mm 0mm, width=0.40\textwidth]{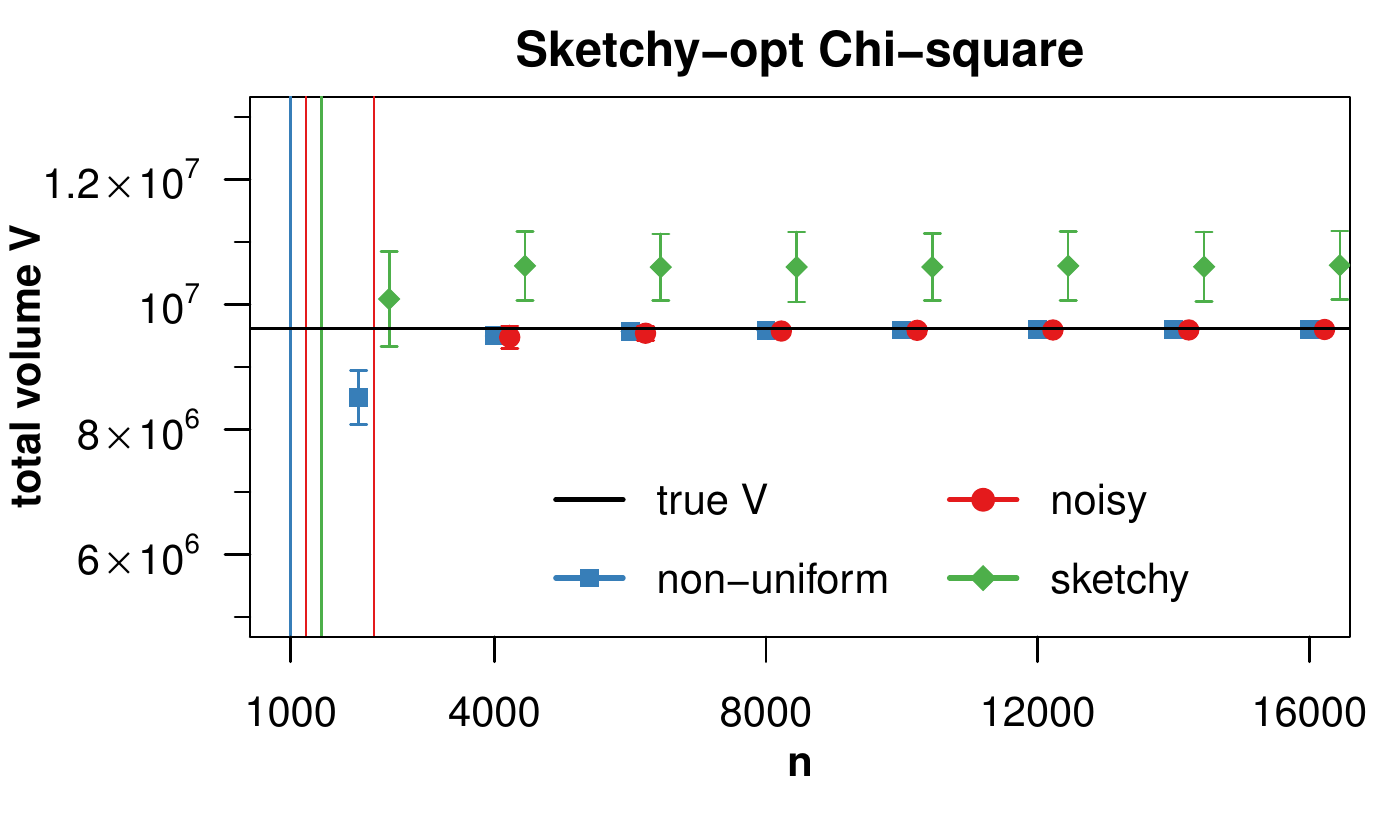}
	\caption{Left: Zipf and Zipf+Sketchy populations. Right: sketchy-optimized volume estimation (binned data).	\label{fig3}}
\end{figure*}

\subsection{Chi-square Minimization}

For binned data, the analogue of Least Square regression method consists of minimizing the $\chi^2$ (Chi-square) statistics~\cite{Berkson1980}.
Let us denote by $n_j^e$ and $n_j^o$ the expected and observed number of queries in the $j$-th bin $[\ell_{j-1},\ell_j)$. The latter is observed from data, while the former can be estimated as follows. From Eq.~\ref{eq:Nv}, the volume $\ell_j$ (resp.,~$\ell_{j-1}$) is the one of the query with rank $\hat{N}_{\ell_j}$ (resp.,~$\hat{N}_{\ell_{j-1}}$).
Thus the expected number of queries in the $j$-th bin is:
$$
n_j^e(c,\beta) = \hat{N}_{\ell_{j-1}} - \hat{N}_{\ell_{j}} = (c/\ell_{j-1})^{1/\beta} - (c/\ell_j)^{1/\beta}  
$$
where we explicitly write the parameters $c$ and $\beta$. 
We now estimate the values of $c$ and $\beta$ as those minimizing the $\chi^2$: 
\begin{equation}\label{eq:chi2}
(\hat c, \hat \beta) = \underset{(c,\beta)}{\operatorname{argmin}} \sum_{j=j_{v_{\mathit{max}}}}^M \frac{\left(n_j^o-n_j^e(c,\beta)\right)^2}{n_j^e(c,\beta)}
\end{equation}
As in the continuous case, we restrict to tail values that best fit a Power-law. This is why the summation in Eq.~\ref{eq:chi2} starts from $j_{v_{\mathit{max}}}$, the bin of the $v_{\mathit{max}}$ value returned by the Binned-CSN method.

Simulation results are shown in Figures~\ref{fig:est:binned:c}--\ref{fig:est:binned:beta}~(right). In particular, the estimation of $\beta$ performs comparably well to the Binned-CSN method for non-uniform and noisy sampling (see Figure~\ref{fig:est:binned:c}). For sktechy sampling, the Chi-square method is preferrable to Binned-CSN. In fact, while it shows a slightly larger variance, it has a significantly smaller bias, summing up to an overall smaller mean square error. 

\subsection{Constrained Chi-square Minimization}

Binned-CSN only provides an estimation for the coefficient $\beta$, while Chi-square provides an estimator for the intercept $c$ as well. We are interested here in the definition of an estimator of $c$ to be used together with Binned-CSN. In the continuous case, the max-estimator performed comparably well to NLS (cfr. Figure~\ref{fig:est:c}). Unfortunately, the maximal observed volume $v_1$ in the sample provides a very biased estimation of $c$ for binned data. This is due to the fact the bin of $v_1$ corresponds to a large range of possible true values. Therefore the estimation of $c$ should take into account the observed values at several bins, as the Chi-square method does. 
We propose to combine the strengths of Binned-CSN and Chi-square methods as follows. First, we obtain an estimate $\hat{\beta}_{\mathit{CSN}}$ of $\beta$ by applying the Binned-CSN method. Then, to estimate $c$ we solve the optimization problem:
\begin{equation}\label{csnBin}
	\hat c =  \underset{c}{\operatorname{argmin}} \sum_{j=j_{v_{\mathit{max}}}}^M \frac{\left(n_j^o-n_j^e(c,\hat{\beta}_{\mathit{CSN}})\right)^2}{n_j^e(c,\hat{\beta}_{\mathit{CSN}})}
\end{equation}
which is a constrained version of Eq.~\ref{eq:chi2}.

Simulations results in Figure~\ref{fig:est:binned:c}~(left) show that Constrained Chi-square provides an almost unbiased estimator of $c$ for non-uniform and noisy sampling, while it under-estimates $c$ for sketchy sampling. Standard deviations are smaller than the Chi-square estimator shown in Figure~\ref{fig:est:binned:c}~(right). However, bias for sketchy sampling is higher.

\subsection{Overall approach} \label{sec:binned:ov}

Assuming no information on the sampling bias, from the previous simulation results, the Chi-square method is to be preferred to Binned-CSN plus Constrained Chi-square for estimating $\hat{\beta}$ and $\hat{c}$. This conclusion is in line with the comparison between CSN and NLS in the continuous case. However, for binned data the difference is not so neat.

Starting from the estimations of $\hat{\beta}$ and $\hat{c}$, the procedures described in Section~\ref{sec:estv} for estimating the total number of queries $N$ in the population and their total volume ${\cal V}$ apply unaltered to the case of binned data. 
Figure~\ref{fig:est:binned:nv} reports the simulation results for the estimation of the population size $N$ and total volume ${\cal V}$. The overall approach performs extremely well for non-uniform and noisy sampling. For the latter, there is even a smaller variance than in the continuous case (cfr. Figures~\ref{fig:est:n}-\ref{fig:est:v} (right)). Intuitively, the variability impact of noise is canceled out when it does not extend beyond the same bin of the true value. Regarding sketchy sampling, instead, the estimator is biased and with non-negligible variance. 

Finally, the procedure of Section~\ref{sec:errors} for calculating errors $\Delta N$ and $\Delta  {\cal V }$ extends to binned data as far as $\Delta \beta$ and $\Delta c$ are provided in input. 
For uniformity with NLS regression, statistical errors $\Delta \beta$ and $\Delta c$ are calculated using an adaption of the linearization method to Chi-square minimization.

\setlength{\tabcolsep}{3.5pt}
\begin{table*}[t]
	\small
	\centering
	\resizebox{\linewidth}{!}{%
	\begin{tabular}{rr|rr|rr|rr|rr}
		& & \multicolumn{4}{c|}{Google Trends} &\multicolumn{4}{c}{SearchVolume}\\ \hline
		$v$ & $v/12$ & $\hat{N}_v$ & $\Delta N_v$ & $\hat{\mathcal V}_v$ & $\Delta {\mathcal V}_v$ & $\hat{N}_v$ & $\Delta N_v$ & $\hat{\mathcal V}_v$ & $\Delta {\mathcal V}_v$\\
		\hline
		12 & 1 & 269,214,520  & $\pm$ 18,507,467 & 14,169.58 M & $\pm$ 827.91 M & 5,849,311,206 & $\pm$ 10,205,374,040 & 157,547.70 M & $\pm$ 262,434.70 M\\
		120 & 10 & 13,770,732 & $\pm$  815,062 & 7,171.15 M & $\pm$  354.03 M & 91,968,610 & $\pm$ 136,211,457 & 24,766.71 M & $\pm$ 34,731.11 M\\
		1,200 & 100 & 704,394  & $\pm$  33,959 & 3,591.35 M & $\pm$  145.85 M & 1,446,021 &  $\pm$ 1,760,408 & 3,889.59 M & $\pm$ 4,433.55 M\\
		12,000 & 1,000 & 36,031 & $\pm$ 1,444 & 1,760.23 M & $\pm$ 56.87 M & 22,736 &  $\pm$ 21,685 & 607.08 M & $\pm$ 535.30 M\\
		120,000 & 10,000 & 1,843 & $\pm$ 56 & 823.63 M & $\pm$ 20.30 M & 357 & $\pm$ 247 & 91.03 M & $\pm$ 58.45 M\\
		600,000 & 50,000 & 231 & $\pm$ 5 & 456.95 M & $\pm$ 9.06 M & 20 & $\pm$ 10 & 21.4 M & $\pm$ 10.89 M\\
		\hline
	\end{tabular}
	}
	\mbox{}\\[2ex]
	\caption{Estimated $N_v$ and ${\mathcal V}_v$ for queries with at least $v$ searches in 2017. $v/12$ is the monthly average of $v$.} 
	\label{tb:estNV}
\end{table*}

\subsection{Sketchy-optimized approach} \label{sec:binned:sk}

Let us finally investigate an approach to reduce bias and variance in the case of sketchy sampling under the \textit{further assumption} that we know that data is actually sketchy sampled and we know the fraction\footnote{Alternatively, $\gamma c$ (which is the actual value needed in the sketchy-optimized approach)
can be estimated starting from a sample of queries for which we know both the true volume $V_i$ and the noisy volume $X_i$. In fact, by Eq.~\ref{eq:sk}, we have that $X_i - V_i = \gamma_i c$ is drawn from a uniform distribution in the range $[0, \gamma c]$.} $\gamma$ in the noise of Eq.~\ref{eq:sk}. The formula in such an equation is a sum of two independent random variables, whose density function can be explicitly calculated using convolution of probability distributions. We omit the details of the calculation, but, as it could be expected, the resulting density is closer to a Zipf for large volumes and it is closer to uniform distribution for small volumes. Figure~\ref{fig3}~(left) shows the population  of a pure Zipf and the effects of adding noise\footnote{Figure~\ref{fig3}~(left) shows the effects on the whole population. The resulting distribution differs (in the tail) from Figure~\ref{fig:simsampling}, where non-uniform sampling is also considered in addition to adding noise.}  as in Eq.~\ref{eq:sk} with $\gamma = 0.001$. Intuitively, volumes smaller than $\gamma c = c/1000$ are heavily modified. Volumes larger than $c/1000$ are less modified. However, the method estimating $\beta$ may understimate it due to the data with volume close to $c/1000$. Around such a volume the Power law starts becoming apparent, but with a biased coefficient. If we concentrate on volumes larger than $10 \gamma c = c/100$, sketchy sampling may impact for at most $10\%$ of the volume, hence reducing the bias due to noise of Eq.~\ref{eq:sk}. 

In summary, we propose the following simple modification to the estimation procedure for sketchy sampling: use sample data whose volume is larger or equal than $10 \gamma v_1$, where $\gamma$ is assumed to be known and $v_1$ is the largest observed volume in the sample data. 

We have experimented this modified procedure for both Binned-CSN and Chi-square methods. The latter continues to perform slightly better even in such setting. Simulation results for the estimation of the total volume are shown in Figure~\ref{fig3}~(right). Contrasted to Figure~\ref{fig:est:binned:nv}~(right), the estimations in the sketchy sampling schema are less biased and with smaller variances. Moreover, such an approach does not degrade the performances of Chi-square for non-uniform and noisy sampling under the simulation parameters. Intuitively, such two sampling schemes do not systematically impact on the slope of the empirical sample distribution, hence restricting to top-volumes (larger or equal than $10 \gamma v_1$) does not affect coefficient estimation.

\section{Empirical analysis}\label{sec:data}

In this section, we conduct an empirical analysis on a real dataset. We generated a sample of 120K queries by crawling 18 popular Italian websites about  recipes and cooking. The list of websites was compiled with the help of web marketing experts and by looking at the rankings of SEO tools\footnote{E.g., \href{https://serpstat.com}{\textit{https://serpstat.com}}}.
Queries were generated in one of the following forms: the name of a recipe as reported in metadata available at web pages\footnote{Such metadata are standardized in the Structured Data format and they are intended to optimize search engine indexing of the web page, see \href{https://developers.google.com/search/docs/data-types/recipe}{\textit{https://developers.google.com/search/docs/data-types/recipe}}.}(e.g.,~``spaghetti with tomato sauce"), queries based on a selected list of ingredients (e.g.,~``recipes with pepperoni"), queries suggested from SEO tools starting from a selected list of keywords, and web marketer expert queries used in past advertising campaigns (in particular, large volume queries such as ~``recipe", ``cake", ``pizza"). Next, queries were validated/filtered by humans as belonging to the recipe and cooking domain. Finally, variants of queries without stop words were also added.
We then submitted the 120K queries to Google Trends (continuous values) and to SearchVolume (binned values) to collect the observed volume of each query for the reference year 2017 and for Italian user agents. Considering a whole year prevents seasonal bias in data. 
Most of the queries received no observed volume, due to the fact they belong to the long tail of the distribution and then SEO tools do not monitor them. This is expected, under the assumption of a Zipf's law distribution of the population of all queries. 

\subsection{Google Trends (continuous data)}\label{sec:gt}

Google Trends has several advantages over other SEO tools. First, the observed volumes provided are computed from (a sample of) the Google search engine query logs, and not from unspecified sources which may have unknown forms of bias. Second, data can be aggregated for arbitrary ranges of time and user agent languages. Most of the other tools, instead, provide monthly averages at the time of request, making it impossible to extend an experiment incrementally to new queries. Third, observed volumes of Google Trends are ratio-scaled, while other SEO tools provide binned values, i.e.,~ranges of observed volumes.
On the negative side, the observed volume provided by Google Trends is relative, not absolute. Google Trends reports the observed volumes of a set of queries in each week of a time period by setting to 100 the largest volume in a week and scaling all other week volumes into the range [0, 100]. We then fixed one specific query to the conventional yearly volume of $1$, and collected observed volume of all other queries in comparison to the specific query. Next, we scaled the relative volumes to  absolute volumes by relying on an estimation procedure of a scaling factor. 
Details on the calculation of relative volumes and of the scaling factor are 
 in~\cite{DBLP:conf/www/LilloR19}.

We obtained observed volumes by Google Trends for about 18.5K queries out of the 120K in the sample set of queries. The resulting rank-volume distribution is shown in  Figure~\ref{fig:empirical}. The remaining queries belong to the long tail, for which Google Trends returns no observed volume.

Let us now apply the estimation model designed in  Section~\ref{sec:model} to the empirical data of Google Trends. 
As shown by the red line fit in Figure~\ref{fig:empirical} (left panel), the NLS regression estimates are:
$$\hat{\beta} = 0.7745\quad  \hat{c} = e^{17.5189} = 40,584,860.$$
The NLS fit is considered for the top $\mathit{max} = 1725$ queries determined by the CSN method.
The statistical errors of the above estimates are considerably low:
\begin{equation}\label{eq:deltagt}
\Delta \beta = 0.0025 \quad \Delta c = 199,263
\end{equation}
We can now use Eq. \ref{eq:N} for estimating the number $N_v$ of queries having a volume of at least $v$, and Eq. \ref{eq:errN} for calculating the statistical error $\Delta N_v$. Similarly, Eq. \ref{est:vv} can be used for estimating the total volume ${\mathcal V}_v$ of queries having a volume of at least $v$, and Eq. \ref{eq:dV} for calculating its statistical error $\Delta {\mathcal V}_v$. Table~\ref{tb:estNV} reports the estimates for a few values of $v$.
As a means of comparison, the total empirical volume of the 18.5K queries in our sample amounts at 1,057 M searches. Such a large number is consistent with the fact that the sample is not uniform, but top ranked queries are more likely to be in the sample. Moreover, it also gives confidence that the sample is sufficiently large (as per empirical volume) to correctly estimate the true volume.
According to the simulations of Figure~\ref{fig:est:v}, the values $\hat{N}_v$ and $\hat{{\mathcal V}}_v$ may overestimate the true $N_v$ and ${\mathcal V}_v$ respectively, if some sketchy approximation is introduced in the observed volume data by Google Trends. 
In case of noisy sampling, instead, a small under or overestimation may occur. It is worth noting that these conclusions hold under the assumption that observed volumes are biased only as modeled in Section~\ref{sec:sampling}. However, other biases may be present in observed volumes from SEO tools.  In the case of Google Trends, one other bias is due to the scaling procedure from relative to absolute volumes.

\subsection{SearchVolume (binned data)}\label{sec:sv}\label{sec:emp:sv}

SearchVolume is a popular SEO tool providing free access to bulk observed volumes. 
Also, SearchVolume provides the observed volume of queries for a few specific countries, including Italy.  The underlying methods and log data used by the system are undisclosed. 
We submitted to SearchVolume the 18.5K queries for which Google Trends provided observed volumes, and obtained (binned) observed volumes for about 12.5K of them. The resulting rank-volume distribution is shown in  Figure~\ref{fig:semrush}~(left). 

A number of facts are worth being pointed out when contrasting that distribution with the one of Google Trends in Figure~\ref{fig:empirical}. First, SearchVolume returned no result for about 6K queries, which are not necessarily low volume ones according to Google Trends. For instance, 26 out of the top 100 Google Trends observed volume queries are assigned no observed volume by SearchVolume. As already pointed out, each SEO tool comes with its own biases on the set of queries covered and on the values of the observed volumes. 
In fact, as a second point, top observed volumes are up to 10$\times$ smaller than the top volumes of Google Trends. 
In particular, the total empirical volume of the 12.5K queries amounts at 206 M searches vs the 805M searches of Google Trends for the same set of queries.
Independent SEO tools are known to be more conservative than Google-owned tools.
Another possible reason is that  the multiplicative factor devised in~\cite{DBLP:conf/www/LilloR19} to scale Google Trends relative volumes may have been over-estimated. Third, Google Trends data has a larger empirical variance than SearchVolume: top volume are up to 65 standard deviations from the mean volume for Google Trends and up to 38 for SearchVolume. 
Fourth, correlation between SearchVolume and Google Trends volumes is weak, with 
Kendall's $\tau = 0.3717$. These last two points could originate by the different sampling biases of the two tools.

The above arguments are not specific of the two tools we are considering in this paper, but they apply when contrasting any pair of SEO tools. In fact, in web marketing practice, the adoption of a specific SEO tool is mainly based on reputation and trust on the tool. 

Let us apply the estimation model designed in  Section~\ref{sec:binned:ov} to the empirical data of SearchVolume. The estimated values are:
$$\hat{\beta} = 0.5545 \quad  \hat{c} = e^{14.9551} = 3,125,485.$$
The Chi-square fit is considered for the top $\mathit{max} = 119$ queries determined by the Binned-CSN method. It turns out that the Sketchy-optimized method of Section~\ref{sec:binned:sk} provided exactly the same results.
The statistical errors of the above estimates are larger than (\ref{eq:deltagt}):
$$\Delta \beta = 0.0352 \quad \Delta c = 549,134$$
This is somehow expected, and it can be attributed to the loss of information due to binning. Numerical experiments on synthetic data shows, in fact, that the error of the coefficients of the Zipf's law are more than one order of magnitude larger than the error when the data are observed without binning.
Finally, the estimated values for the number $N_v$ of queries having a volume of at least $v$, and the  total volume ${\mathcal V}_v$ are reported in Table~\ref{tb:estNV}. 

Contrasting the estimations inferred from SearchVolume and Google Trends data, the smaller coefficient of the former implies a larger number of queries in the long tail. There are, in fact, 5.85B queries searched at least once a month for SearchVolume vs only about 270M for Google Trends. On the contrary, a larger number of top-volume queries are estimated from Google Trends data. The two methods get closer in estimating the number of queries with at least 1,000 searches per month (23K vs 36K), and in estimating the total volume of queries with at least 100 searches per month (3.9B vs 3.6B). Finally, the larger statistical errors $\Delta \beta$ and $\Delta c$ produce considerably larger estimates $\Delta N_v$ and $\Delta {\mathcal V}_v$ for SearchVolume in comparison to Google Trends.

\section{Conclusion}

We investigated the problem of estimating the total number of searches of queries belonging to a specific domain in a given period of time. By doing the sensible assumption that the unobserved rank distribution of query volumes follows a Zipf's law, our approach can be decomposed in two parts. First, we model biases in obtaining observed volumes from SEO tools. Such biases consist of non-uniform sampling possibly coupled with noise and approximation errors. Second, we devised estimation methods to infer the total volume of the queries of the domain starting from a biased sample. The estimation methods distinguish continuous and binned empirical data. They are able to find the total number and the total volume of the queries in the domain which have been searched at least $v$ times in a given time period. This kind of information is extremely useful in web marketing research and advertising to quantify the market value of a domain. A large set of numerical simulations supports the validity of the proposed methods. Finally, we presented an empirical application w.r.t.~the domain of \textit{recipes and cooking} for Italian searches in 2017, including a comparison of the continuous vs binned data cases. 

The \textit{first critical issue} for extending our analysis to other domains consists of checking the hypothesis that the population of queries in the domain is Zipfian.
As shown in Figures~\ref{fig:empirical} and~\ref{fig:semrush}~(left), empirical data in the domain of recipes and cooking appear to be Zipfian. This motivated our assumption that the reference population, namely the queries searched in a reference domain, follows a Zipf's law.  Ref. \cite{Cristelli2012} points out that the granularity and extent of a reference population  should exhibit a ``coherence" property. This is particularly relevant, since splitting or merging two Zipfian sets does not necessarily yield another Zipfian set, hence the actual definition of what is and what is not in a domain is essential in meeting our assumption.
The domain considered in this paper has well-defined boundaries that make it reasonably coherent. 

The \textit{second critical issue} is the construction of the sample set of queries. As shown by the numerical simulations, the capability of correctly inferring the total volume significantly depends on the inclusion in the empirical sample of top volume queries from the population. 
Empirical sample queries should then carefully collected, e.g.,~by resorting to domain's expert knowledge or, if feasible, by crawling a set of specialized websites. Finding estimators which are (more) robust to the choice of the query sample is certainly an interesting potential extension of our approach to the case when it is costly to construct sufficiently large samples. 

The \textit{third critical issue} is concerned with understanding (eg.,~through statistical tests) which type of bias is likely to be present in empirical data provided by a specific SEO tool. In this paper, we considered three possible scenarios: uniform sampling alone, or together with normally distributed noise (noisy sampling), or together with count-min sketch like approximation (sketchy sampling). Other scenarios can be conceived, e.g.,~noise due to data anonymization~\cite{DBLP:conf/ndss/MelisDC16,DBLP:conf/icde/CormodeKS18}. Further work is necessary to test which scenario fits better for a given empirical data.

Finally, the estimation methods devised in this paper are generally applicable to any context where the population is Zipfian. One such context, suggested by an anonymous reviewer, regards the Internet Domain Name System (DNS). Here, client machines query a distributed database for resolving the numeric IP address of a given host name. A DNS server local to an organization may maintain (approximate) counts of the number of queries per host name. The volume per host name is known to follow a Zipf's law~\cite{DBLP:journals/ton/JungSBM02}. Hence, our methods can be used to estimate the total number of distinct host names (in our notation, $N$) served by a local DNS server without having to store them explicitly -- which could compromise efficiency of the DNS server. Also, by aggregating the counts of top volume host names for several DNS servers, we can estimate the size of the Internet (the number of IP addresses) accessed by the user base served.

\section*{Software code}

Software code in R \cite{Rlang} 
of all estimation methods is available at \href{https://github.com/ruggieris/QVolume}{\textit{https://github.com/ruggieris/QVolume}}.

\ifCLASSOPTIONcompsoc
  \section*{Acknowledgments}
\else
  \section*{Acknowledgment}
\fi

This work has been partially supported by a research grant by \textit{ForTop S.R.L.} (\href{https://www.fortop.it/en}{\textit{https://www.fortop.it/en}}) on the topic: \textit{Data-driven analysis of search engine query market}. We are  grateful to Lorenzo Barsotti, Stefania Camarda, Paolo Ferragina, Riccardo Guidotti, Marco Marino, and Anna Monreale for stimulating discussions.

\ifCLASSOPTIONcaptionsoff
  \newpage
\fi



\bibliographystyle{IEEEtran}

\bibliography{biblio}

\begin{thebibliography}{10}
\providecommand{\url}[1]{#1}
\csname url@samestyle\endcsname
\providecommand{\newblock}{\relax}
\providecommand{\bibinfo}[2]{#2}
\providecommand{\BIBentrySTDinterwordspacing}{\spaceskip=0pt\relax}
\providecommand{\BIBentryALTinterwordstretchfactor}{4}
\providecommand{\BIBentryALTinterwordspacing}{\spaceskip=\fontdimen2\font plus
\BIBentryALTinterwordstretchfactor\fontdimen3\font minus
  \fontdimen4\font\relax}
\providecommand{\BIBforeignlanguage}[2]{{%
\expandafter\ifx\csname l@#1\endcsname\relax
\typeout{** WARNING: IEEEtran.bst: No hyphenation pattern has been}%
\typeout{** loaded for the language `#1'. Using the pattern for}%
\typeout{** the default language instead.}%
\else
\language=\csname l@#1\endcsname
\fi
#2}}
\providecommand{\BIBdecl}{\relax}
\BIBdecl

\bibitem{DBLP:journals/tois/PetersenSL16}
C.~Petersen, J.~G. Simonsen, and C.~Lioma, ``Power law distributions in
  information retrieval,'' \emph{{ACM} Trans. Inf. Syst.}, vol.~34, no.~2, pp.
  8:1--8:37, 2016.

\bibitem{DBLP:journals/ftdb/CormodeGHJ12}
G.~Cormode, M.~N. Garofalakis, P.~J. Haas, and C.~Jermaine, ``Synopses for
  massive data: Samples, histograms, wavelets, sketches,'' \emph{Foundations
  and Trends in Databases}, vol.~4, no. 1-3, pp. 1--294, 2012.

\bibitem{DBLP:journals/jal/CormodeM05}
G.~Cormode and S.~Muthukrishnan, ``An improved data stream summary: the
  count-min sketch and its applications,'' \emph{J. Algorithms}, vol.~55,
  no.~1, pp. 58--75, 2005.

\bibitem{DBLP:journals/siamrev/ClausetSN09}
A.~Clauset, C.~R. Shalizi, and M.~E.~J. Newman, ``Power-law distributions in
  empirical data,'' \emph{{SIAM} Review}, vol.~51, no.~4, pp. 661--703, 2009.

\bibitem{VirkarClauset2014}
Y.~Virkar and A.~Clauset, ``Power-law distributions in binned empirical data,''
  \emph{The Annals of Applied Statistics}, vol.~8, no.~1, pp. 89--119, 2014.

\bibitem{doi:10.1080/00107510500052444}
M.~Newman, ``Power laws, pareto distributions and {Z}ipf's law,''
  \emph{Contemporary Physics}, vol.~46, no.~5, pp. 323--351, 2005.

\bibitem{DBLP:conf/wsdm/DingABS11}
S.~Ding, J.~Attenberg, R.~A. Baeza{-}Yates, and T.~Suel, ``Batch query
  processing for web search engines,'' in \emph{{WSDM}}.\hskip 1em plus 0.5em
  minus 0.4em\relax {ACM}, 2011, pp. 137--146.

\bibitem{DBLP:conf/kdd/Baeza-YatesT07}
R.~A. Baeza{-}Yates and A.~Tiberi, ``Extracting semantic relations from query
  logs,'' in \emph{{KDD}}.\hskip 1em plus 0.5em minus 0.4em\relax {ACM}, 2007,
  pp. 76--85.

\bibitem{DBLP:conf/sigir/Baeza-YatesGJMPS07}
R.~A. Baeza{-}Yates, A.~Gionis, F.~Junqueira, V.~Murdock, V.~Plachouras, and
  F.~Silvestri, ``The impact of caching on search engines,'' in
  \emph{{SIGIR}}.\hskip 1em plus 0.5em minus 0.4em\relax {ACM}, 2007, pp.
  183--190.

\bibitem{DBLP:journals/jasis/Bookstein90}
A.~Bookstein, ``Informetric distributions, part {I:} unified overview,''
  \emph{{JASIS}}, vol.~41, no.~5, pp. 368--375, 1990.

\bibitem{Adamic02}
L.~Adamic and B.~Huberman, ``Zipf's law and the internet,''
  \emph{Glottometrics}, vol.~3, pp. 143--150, 2002.

\bibitem{epjb}
M.~Goldstein, S.~Morris, and G.~Yena, ``Problems with fitting to the power-law
  distribution,'' \emph{European Physical Journal B}, vol.~41, pp. 255--258,
  2004.

\bibitem{JSSv064i02}
C.~Gillespie, ``Fitting heavy tailed distributions: The powerlaw package,''
  \emph{J. of Stat. Software}, vol.~64, no.~2, pp. 1--16, 2015.

\bibitem{Rlang}
\BIBentryALTinterwordspacing
{R Core Team}, \emph{R: A Language and Environment for Statistical Computing},
  R Foundation for Statistical Computing, Vienna, Austria, 2019. [Online].
  Available: \url{https://www.R-project.org/}
\BIBentrySTDinterwordspacing

\bibitem{NBERt0342}
X.~Gabaix and R.~Ibragimov, ``Rank — 1/2: A simple way to improve the {OLS}
  estimation of tail exponents,'' \emph{Journal of Business \& Economic
  Statistics}, vol.~29, no.~1, pp. 24--39, 2011.

\bibitem{RePEc:gla:glaewp:2012_10}
G.~Fazio and M.~Modica, ``Pareto or log-normal? best fit and truncation in the
  distribution of all cities,'' \emph{Journal of Regional Science}, vol.~55,
  no.~5, pp. 736--756, 2015.

\bibitem{Clauset2019}
A.~D. Broido and A.~Clauset, ``Scale-free networks are rare,'' \emph{Nature
  Communications}, vol.~10, 2019.

\bibitem{Holme2019}
P.~Holme, ``Rare and everywhere: {P}erspectives on scale-free networks,''
  \emph{Nature Communications}, vol.~10, p. 1016, 2019.

\bibitem{DBLP:journals/corr/abs-1811-02071}
I.~Voitalov, P.~van~der Hoorn, R.~van~der Hofstad, and D.~Krioukov,
  ``Scale-free networks well done,'' \emph{Phys. Rev. Research}, vol.~1, p.
  033034, 2019.

\bibitem{pnas2016}
A.~Orlitskya, A.~Sureshb, and Y.~Wuc, ``Optimal prediction of the number of
  unseen species,'' \emph{Proceedings of the National Academy of Sciences USA},
  vol. 113, pp. 13\,283--13\,288, 2016.

\bibitem{DBLP:journals/jasis/VaughanC15}
L.~Vaughan and Y.~Chen, ``Data mining from web search queries: {A} comparison
  of {G}oogle {T}rends and {B}aidu {I}ndex,'' \emph{{JASIST}}, vol.~66, no.~1,
  pp. 13--22, 2015.

\bibitem{DBLP:conf/www/LilloR19}
F.~Lillo and S.~Ruggieri, ``Estimating the total volume of queries to
  {G}oogle,'' in \emph{{WWW}}.\hskip 1em plus 0.5em minus 0.4em\relax {ACM},
  2019, pp. 1051--1060.

\bibitem{Berkson1980}
J.~Berkson, ``Minimum chi-square, not maximum likelihood!'' \emph{The Annals of
  Statistics}, vol.~8, pp. 457--487, 1980.

\bibitem{Cristelli2012}
M.~Cristelli, M.~Batty, and L.~Pietronero, ``There is more than a power law in
  {Z}ipf,'' \emph{Scientific Reports}, vol.~2, p. 812, 2012.

\bibitem{DBLP:conf/ndss/MelisDC16}
L.~Melis, G.~Danezis, and E.~{De Cristofaro}, ``Efficient private statistics
  with succinct sketches,'' in \emph{{NDSS}}.\hskip 1em plus 0.5em minus
  0.4em\relax The Internet Society, 2016.

\bibitem{DBLP:conf/icde/CormodeKS18}
G.~Cormode, T.~Kulkarni, and D.~Srivastava, ``Constrained private mechanisms
  for count data,'' in \emph{{ICDE}}.\hskip 1em plus 0.5em minus 0.4em\relax
  {IEEE}, 2018, pp. 845--856.

\bibitem{DBLP:journals/ton/JungSBM02}
J.~Jung, E.~Sit, H.~Balakrishnan, and R.~T. Morris, ``{DNS} performance and the
  effectiveness of caching,'' \emph{{IEEE/ACM} Trans. Netw.}, vol.~10, no.~5,
  pp. 589--603, 2002.

\end{thebibliography}
\begin{IEEEbiography}[{\includegraphics[width=1in,height=1.25in,clip,keepaspectratio]{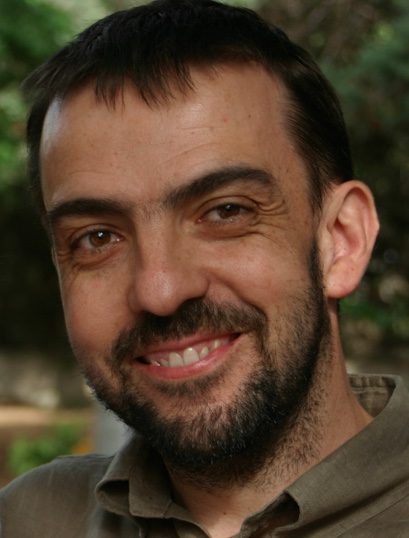}}]{Fabrizio Lillo,}Ph.D. is professor of Mathematical Methods for Economics and Finance at the University of Bologna (Italy), teaching at the programs of Quantitative Finance and Mathematical Finance at the M.Sc and Ph.D. levels. He is a member of faculty board of the Ph.D. program in Data Science of the Scuola Normale Superiore in Pisa. His research interests focus on statistical methods and mathematical models for finance, economics, and social sciences.
\end{IEEEbiography}
\begin{IEEEbiography}[{\includegraphics[width=1in,height=1.25in,clip,keepaspectratio]{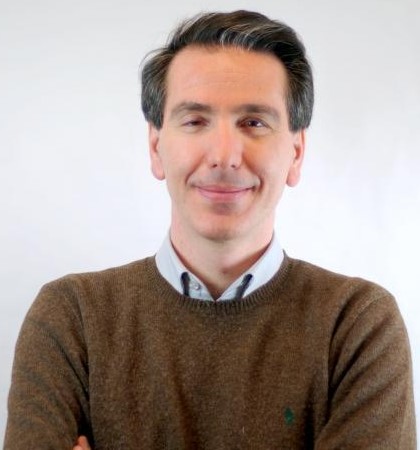}}]{Salvatore Ruggieri,}Ph.D.  is professor of Computer Science at the University of Pisa, where he teaches Data Science and Web Marketing. He holds a PhD in Computer Science (1999). He is a member of the KDDLab (\href{http://kdd.isti.cnr.it}{\textit{http://kdd.isti.cnr.it}}), with research interests in  algorithmic fairness, explainable AI, privacy, modelling the process of knowledge discovery, classification algorithms, web mining, and applications.
\end{IEEEbiography}
\end{document}